\documentclass{iopart}
\usepackage{iopams,amsmath,bm,algorithm,algorithmic}
\usepackage{array,eqparbox,dcolumn,multirow,slashbox}
\usepackage[pdftex]{graphicx,color}
\graphicspath{{figures/pdf/},{figures/jpg/}}
\DeclareGraphicsExtensions{.pdf}
\usepackage{subfig}
\def\norm#1{\sigma \left( #1 \right)}
\def\vec#1{\mathbf{#1}}

\def\Re{\mathrm{Re}}

\def\norm#1{\| #1 \|}

\def\Tr{\operatorname{Tr}}

\def\F{\mathcal{F}}

\def\D{\mathcal{D}}
\def\E{\mathcal{E}}
\def\L{\mathcal{L}}
\def\ONE{\mathbb{I}}
\def\tp{\mathrm{T}}
\def\tf{T}

\begin{document}

\title[Robust Quantum Gates]{Robust quantum gates for open systems via
optimal control: Markovian versus non-Markovian dynamics}

\author{Frederik F Floether}

\address{Dept of Physics, Cavendish Laboratory, University 
of Cambridge, J. J. Thomson Avenue, Cambridge, CB3 0HE, United Kingdom}

\author{Pierre de Fouquieres}

\address{Dept of Applied Mathematics and Theoretical Physics, 
 University of Cambridge, Wilberforce Road, Cambridge, CB3 0WA, United
 Kingdom}

\author{Sophie G Schirmer}
               
\address{Dept of Physics, College of Science, Swansea University,
Singleton Park, Swansea, SA2 8PP, United Kingdom \\[1ex]

Dept of Applied Mathematics and Theoretical Physics, University of Cambridge, \\
Wilberforce Road, Cambridge, CB3 0WA, United Kingdom}

\ead{fff22@cam.ac.uk, sgs29@swan.ac.uk}

\begin{abstract}
We study the implementation of one-, two-, and three-qubit quantum gates
for interacting qubits using optimal control.  Different Markovian and
non-Markovian environments are compared and efficient optimisation
algorithms utilising analytic gradient expressions and quasi-Newton
updates are given for both cases.  The performance of the algorithms is
analysed for a large set of problems in terms of the fidelities attained
and the observed convergence behaviour.  New notions of success rate and
success speed are introduced and density plots are utilised to study the
effect of key parameters, such as gate operation times, and random
variables, such as the initial fields required to start the iterative
algorithm.  Core characteristics of the optimal fields are statistically
analysed.  Substantial differences between Markovian and non-Markovian
environments in terms of the possibilities for control and the control
mechanisms are uncovered.  In particular, in the Markovian case it is
found that the optimal fields obtained without considering the
environment cannot be improved substantially by taking the environment
into account and the fidelities attained are determined mostly
by the gate operation time as well as the overall strength of the environmental
effects.  Computation time is saved if the fields are pre-optimised
neglecting decoherence.  In the non-Markovian case, on the other hand,
substantial improvements in the fidelities are observed when the details
of the system-bath coupling are taken into account.  In that case,
field leakage is shown to be a significant issue which can make high gate
fidelities impossible to obtain unless both the system and noise
qubits are fully controlled.
\end{abstract}

\maketitle

\tableofcontents

\section{Introduction}

After the initial focus on the theoretical basis of quantum mechanics,
research into harnessing the non-classical properties of quantum systems
for technological purposes has intensified over the last decades.  While
some ideas such as quantum key distribution have become commercially
viable applications, the path towards scalable quantum simulation and
computation has turned out to be stony.  One of the highest hurdles
has been the influence of the environment on quantum systems, causing
decoherence and eliminating purely quantum properties~\cite{1}.
Multiple strategies to mitigate against the effects of decoherence have
been proposed; for example, decoherence-free subspaces and noiseless
subsystems~\cite{2}, quantum dynamical decoupling~\cite{3}, multilevel
encoding of logical states~\cite{4}, and stochastic and optimal
control~\cite{5}.  In this work, we focus on the last approach.

The basic premise behind optimal control of quantum dynamics is that
quantum-mechanical systems can coherently interact with suitable
external fields, which can change the system's Hamiltonian and thereby
alter its dynamical evolution.  Varying the temporal, and in some cases
spatial profile, of the fields affords us a great degree of control over
the system by enabling us to fine-tune its Hamiltonian at least within
certain constraints.  The idea of optimal control is to use this degree
of control to optimally steer the quantum dynamics towards a desired
outcome, for instance, from a particular initial state to a desired
target state (which might optimise a certain property of the system such
as its energy, dipole moment, or angular momentum).  More recently, with
the advent of quantum information processing, a more ambitious goal has
been formulated: Quantum process control, which aspires to control not
only the evolution of a single initial state but that of a complete set
of basis states to implement a desired dynamic transformation often
referred to as a quantum gate.  Optimal steering of quantum dynamics is
generally considered crucial to achieve high-fidelity implementations of
quantum gates, especially in the presence of environmental noise;
previous work has shown that optimal control can be an effective
strategy to implement quantum gates in noisy settings~\cite{6,44}.
However, it is also known that Hamiltonian control of the system alone
cannot undo environmental effects, especially in the Markovian
setting~\cite{allan}, showing that there are limits on what can be
achieved with optimal control.  Some of these restrictions can be
overcome by extending the idea of control to include measurements or
other forms of incoherent control and feedback, as discussed
in~\cite{allan}, for example. However, we shall focus on open-loop
coherent control, the most widely used technique today.

The main focus of this paper is the potential and limitations of optimal
control with regard to the implementation of high-fidelity quantum gates
in noisy environments with a particular emphasis on the differences
between Markovian and non-Markovian environments.  From a control point
of view these environments are fundamentally different. In the Markovian
case, the system's future depends only on its present state and any
information leaked into the environment is irrecoverable. In the
non-Markovian case, the environment displays memory effects, which the
control can exploit to recover losses to the environment and restore the
system's coherence.  We therefore expect the control mechanisms and
optimal controls for systems in Markovian and non-Markovian environments
to be fundamentally different.  In both cases, however, finding controls
that achieve the desired optimal steering is an optimisation problem
that quickly becomes computationally very expensive as the Hilbert space
dimension of the system increases, due to the need to simulate the
underlying quantum dynamics.  Therefore, the ability of any particular
algorithm to \emph{efficiently} find solutions close to a \emph{global}
optimum is of utmost interest, as are the physical characteristics of
the optimal controls found, the basic mechanisms by which they achieve
the optimal steering, and the effects that may interfere with the
control, potentially rendering it ineffective.  These are the core
issues we shall discuss.  The paper is organised as follows.  In Sec.~2,
we briefly describe the different types of environments and how we model
them.  In Sec.~3, we discuss the optimisation algorithms. The results
for both models are presented and discussed in Sec.~4 and the
conclusions and suggestions for future work are summarised in Sec.~5.

\section{Models of the Environment}

Two idealised environmental models can be distinguished, the Markovian
and the non-Markovian one.  In the former case, the system of interest,
or system for short, interacts with a typically much larger memoryless
environment; it is assumed that the future evolution of the system is
determined solely by its present state.  In addition, the noise signal
has no self-correlation over any time interval.  These conditions are
not fulfilled in the non-Markovian model and the future evolution of the
system also depends on its past.  This is typically true for small
structured environments with a low number of degrees of freedom which
cause the noise to be self-correlated.  As no information is lost, the
coherence of the system can oscillate. Markovian and non-Markovian
settings represent opposite ends of the spectrum and many physical
systems have features in-between the two extremes. It is also possible
for the combination of a system with its non-Markovian environment to
itself be coupled to a Markovian bath, as considered in \cite{44}.  As
such, studying the behaviour of a system under both scenarios allows a
wide range of potential implementations to be covered.

\subsection{Markovian Environments}

The key assumption underlying the Markovian model is that the timescale
relevant for the system evolution is much greater than the coherence
time of the bath so that on the relevant timescales the environment has
no memory of the system's past.  The description of a large memoryless
environment is a very natural one and many systems have traditionally
been studied using the Markovian model.  Among the most common physical
processes responsible for Markovian decoherence are phase and population
relaxation processes.  Dephasing can be caused by external fields or
collisions in atomic vapours~\cite{9}. Population relaxation typically
occurs as a result of spontaneous emission of photons or phonons, which
been shown to play a dominant role in trapped ions~\cite{10}.  Markovian
decoherence processes have also been found in Bose-Einstein condensates
and plasmas~\cite{11,12}.

The dynamics of a system in a Markovian environment are usually
described by a master equation for the reduced density matrix $\rho_r$
obtained by tracing out the bath degrees of freedom from the state of
the system and environment.  The most general time-homogeneous Markovian
master equation, which is also trace-preserving and completely positive
for any initial condition, is the master equation in Lindblad
form~\cite{7}
\begin{equation}
  \dot{\rho}_r(t) = (\L_0 + \L_c + \L_D) \rho_r(t) ,
\end{equation}
where $\L_0\rho_r(t)=[-iH_0,\rho_r(t)]$ describes the action of the
system's intrinsic Hamiltonian $H_{0}$, $\L_c\rho_r(t)
=[-iH_c,\rho_r(t)]$ encapsulates the effect of the control Hamiltonian
$H_{c}$, and $\L_D\rho_r(t)=\sum_d \D[V_d]\rho_r(t)$ accounts for the
environment.  $[A,B]=AB-BA$ is the usual matrix commutator.  The sum
over $d$ represents the sum over all the bounded Lindblad operators
$V_{d}$ acting on the system, where
\begin{equation}
  \D[V_d] \rho_r(t) = V_d \rho_r(t) V_d^\dag 
   - \tfrac{1}{2} [V_d^\dag V_d \rho_r(t)+\rho_r(t)V_d^\dag V_d].
\end{equation}
We choose units such that $\hbar=1$.  Instead of describing the dynamics
only for a specific state $\rho(t)$, we can introduce a superoperator
$X(t)$ describing the evolution of all possible system states, which
must satisfy
\begin{equation}
  \tfrac{d}{dt}X(t) = (\L_0+\L_c+\L_D)X(t), \quad X(0)=\ONE.
\end{equation}

Denoting the system's Hilbert space dimension by $N_1$, the evolution
operator $\L_0 + \L_c + \L_D$ is a linear map between $N_1 \times N_1$
complex matrices and thus can be written as a $N_1^2 \times N_1^2$
complex matrix acting on a complex column vector representing the state.
The latter can be obtained simply by stacking the columns of the density
matrix $\rho$.  An alternative representation, which is often
computationally more convenient, can be obtained by expanding the state
and the evolution operators with respect to a basis
$\{\sigma_k\}_{k=1}^{N_1^2}$ for the Hermitian $N_1\times N_1$ matrices,
such as the generalised Pauli matrices or tensor products of Pauli
matrices for qubit systems.  In this case, the state is represented by a
real $N_1^2$ vector $\vec{r}=(r_n)$, which is determined by $\rho=\sum_n
r_n \sigma_n$, as well as $\L_0$, $\L_c$, and $\L_D$ which are
determined by $N_1^2\times N_1^2$ real matrices $(L_{mn}^{(0)})$,
$(L_{mn}^{(c)}),$ and $(D_{mn}^{(d)})$, respectively~\cite{8,Lendi};
\begin{subequations}
 \begin{align}
   L_{mn}^{(0)} &= \Tr(iH_0 [\sigma_m,\sigma_n]), \\
   L_{mn}^{(c)} &= \Tr(iH_c [\sigma_m,\sigma_n]),  \\
   D_{mn}^{(d)} &= \Tr(V_d^\dag \sigma_m V_d \sigma_n)
   -\tfrac{1}{2} \Tr [V_d^\dag V_d (\sigma_m \sigma_n + \sigma_n \sigma_m)].
\end{align}
\end{subequations}
Choosing a real representation is preferable as the multiplication of
two complex quantities requires four (real) multiplications, compared to
one for two real quantities.

\subsection{Non-Markovian Environments}

Although Markovian environments play an important role in many physical
applications, the study of non-Markovian systems, particularly in the
solid state, has intensified in recent years; non-Markovian behaviour
has been observed in the fluorophore system~\cite{16}, spin
gases~\cite{17}, spin echoes~\cite{18}, quantum dots~\cite{19}, and
donor-based charge qubits interacting with phonons~\cite{20}. Coherence
revivals, a key signature of non-Markovian behaviour, have also been
observed with atoms interacting with cavity electromagnetic radiation
\cite{21} and with electromagnetically trapped ions and molecules
\cite{22,23,24}.

\begin{figure*}
\includegraphics[width=\textwidth]{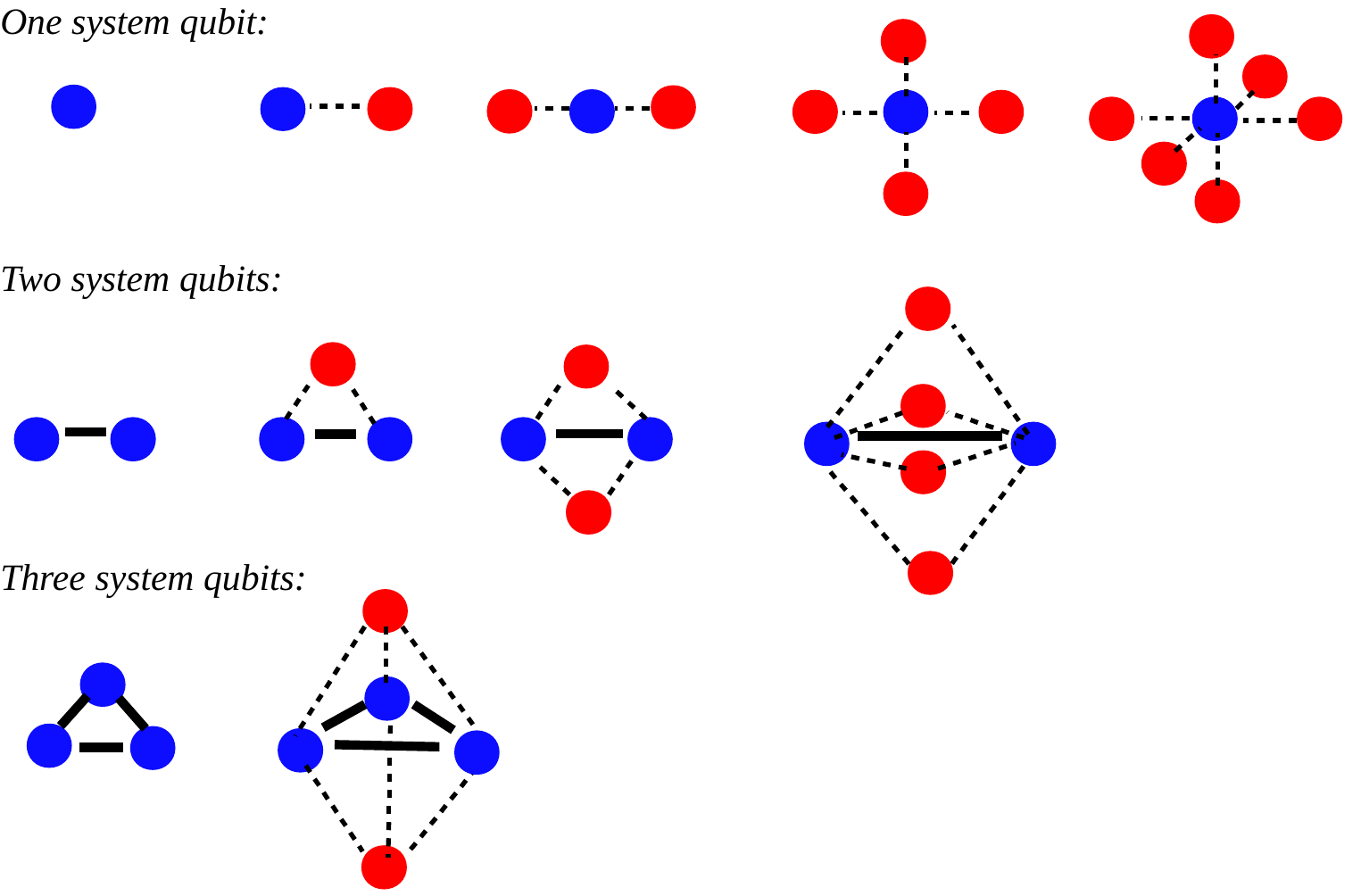} 
\caption{Configurations of system (blue) and noise (red) qubits studied
in the non-Markovian case. Couplings between the noise qubits are
neglected and system-system qubit couplings are assumed to be stronger
than system-noise qubit couplings.}  \label{fig:systems}
\end{figure*}

There are several ways of dealing with non-Markovian environments.  One
approach is to derive master equations similar to the Lindblad equation
for quantum systems under the influence of non-Markovian environments
using perturbative techniques~\cite{13,hwang,schmidt}.  This approach
has the advantage that the evolution of the system is effectively still
determined only by a reduced density matrix describing the system state,
not the state of the environment.  It is very useful to describe large
environments especially if the non-Markovian effects are not too strong.
Another way to deal with environments are collisional model~\cite{14}.
The most general approach are Hamiltonian models that describe the
dynamics of both system and environment and their interaction.  These
models are computationally very challenging, and therefore most useful
small environments, but have the advantage of being able to describe
arbitrary system-bath interactions.  They are therefore useful to model
the dynamics if the system and environment are too intertwined to
describe the system evolution by a master equation.  Here we focus on
the latter full Hamiltonian approach and model the environment in terms
of a small number of noise qubits interacting with one or more system
qubits.  We consider different configurations of system and noise qubits
as shown in Fig.~\ref{fig:systems} similar to \cite{6}.  The coupling
between system qubits as well as system and noise qubits is given by a
Heisenberg exchange interaction and magnetic control fields act on the
system qubits.  The coupling between system qubits is assumed to be
stronger than the coupling between system and noise qubits.  Couplings
between the noise qubits are neglected --- they have been shown to have
little effect on the solutions~\cite{15}.

For a system of $n_1$ system qubits, $n_2$ noise qubits, and $M$ control
Hamiltonians, the dynamics of the composite system are therefore governed
by the Schr\"{o}dinger equation with an overall Hamiltonian 
\begin{equation}
  H_{tot}(t) = \sum_{k=1}^{n_1+n_2} \omega_k S_k^z 
  + \sum_{i=1}^{n_1+n_2-1} \sum_{j>i} \gamma_{ij} \vec{S}_i \cdot
  \vec{S}_j + \sum_{m=1}^M f_m(t) H_m,
\end{equation}
where $N_2=2^{n_2}$ and $N=2^{n_1+n_2}$ are the dimensions of the
environment and the composite system, respectively, and $N_1$ is the
(Hilbert space) dimension of the system as in the Markovian case.
$\vec{S}_i=(S_i^x,S_i^y,S_i^z)$ denotes the spin operators for the $i$th
qubit, where $S_i^x$ is an $n$-fold tensor product whose $i$th factor is
$\frac{1}{2}\sigma_x$ and all other factors are the identity $\ONE$.
$\gamma_{ij}$ is the coupling constant between qubits $i$ and $j$.

\section{Optimal Control Theory \& Algorithms}

\subsection{Discretisation and Optimisation Algorithm}

We use model-based optimal control where an objective functional is
defined and the control field is numerically optimised to obtain the
best possible value of the functional.  A canonical choice for the
performance index for gate optimisation problems is the gate error $\E$.
\footnote{We do not add penalty terms.  Although a popular choice, they
generally complicate the problem and are unnecessary or even
undesirable, especially when dealing with finite time
resolutions~\cite{25}.}
We discretise the controls $f_m(t)$ using piecewise constant functions,
dividing the total time $\tf$ into $K$ slices $[t_{p-1},t_p]$ of
duration $\Delta t=\tf/K$ and setting $f_m(t)=f_{mp}$ for $t \in
[t_{p-1}, t_p]$.  Thus, we have
\begin{equation}
  \label{eq:f}
  f_m(t)=\sum_{p=1}^K f_{mp} \chi_p(t),
\end{equation}
where $\chi_p(t)=1$ for $t\in [t_{p-1},t_p]$ and $0$ elsewhere.  This
parameterisation is a common choice and natural for many applications
such as nuclear magnetic resonance, where waveform generators can
approximate a piecewise constant output.  Other parameterisations may be
considered as appropriate for other applications.  After discretisation
of the controls, we are left with a vector of control parameters
$\vec{f}$ and an error functional $\E(\vec{f})$ and our goal is to find
a control vector $\vec{f}_*$ such that $\E(\vec{f}_*)$ assumes its
global minimum.

Motivated by a recent comparison of various first- and second-order
sequential and concurrent update methods~\cite{29}, we opt for a
concurrent update quasi-Newton method%
~\footnote{This method was chosen as it is generally effective and
efficient; in particular, it outperformed Krotov-type methods with
sequential updates at high fidelities, the regime of interest here.}
, which involves iteratively updating the control parameters according
to the Newton update rule,
\begin{equation}
  \vec{f}^{(k+1)} = \vec{f}^{(k)} - \alpha_k(\mathbf{H}_k)^{-1} \nabla \E(\vec{f}^{(k)}) ,
\end{equation}
starting with a guess for the initial field $f^{(0)}$ and an initial
Hessian approximation $\mathbf{H}_0$.  The initial Hessian is taken to
be the identity and the approximate Hessian is then constructed from the
past gradient history according to the Broyden-Fletcher-Goldfarb-Shanno
(BFGS) formula
\begin{equation}
  (\mathbf{H}_k)^{-1} = 
  \left(1-\frac{\vec{g}_{k-1} \vec{s}_{k-1}^\tp}{\vec{g}_{k-1}^\tp \vec{s}_{k-1}}\right)
  (\mathbf{H}_{k-1})^{-1} 
 \left(1-\frac{\vec{g}_{k-1} \vec{s}_{k-1}^\tp}{\vec{g}_{k-1}^\tp
  \vec{s}_{k-1}}\right)
  + \frac{\vec{s}_{k-1} \vec{s}_{k-1}^\tp}{\vec{g}_{k-1}^\tp \vec{s}_{k-1}} ,
\end{equation}
where $\vec{s}_{k-1}=\vec{f}^{(k)}-\vec{f}^{(k-1)}$, $\vec{g}_{k-1}=\nabla
\E(\vec{f}^{(k)})-\nabla \E(\vec{f}^{(k-1)})$, and $\alpha_k = 1$ by default; $\alpha_k$ is the search
length parameter for a standard Newton step.

The optimisation algorithm requires the computation of derivatives of
the objective functional $\E=\E(f_{mp})$ with regard to the control
variables $f_{mk}$.  The simplest approach is to approximate the required
derivatives numerically using finite differences~\cite{25},
\begin{equation}
  \frac{\partial \E(f_{mp})}{\partial f_{mp}}
 = \lim_{\Delta f_{mp}\to 0} \left\{\frac{1}{\Delta f_{mp}}
   \left[\E(f_{mp}+\Delta f_{mp})-\E(f_{mp})\right] \right\},
\end{equation}
with
\begin{equation}
  \frac{\partial \E(f_{mp})}{\partial f_{mp}}
  \approx \frac{1}{\Delta f_{mp}}
          \left[\E(f_{mp}+\Delta f_{mp})-\E(f_{mp})\right],
\end{equation}
for a finite value of $\Delta f_{mp}$.  This calculation can be made
efficient by recognising that the error does not need to be explicitly
evaluated at the two different fidelities~\cite{26}.  A serious drawback,
however, is that the value of the step size is constrained by the fact
that choosing too large a value renders the approximation invalid while
choosing too small a value reduces the number of accurate digits of the
difference in the floating point representation to zero~\cite{26}.  The
step-size parameter therefore needs to be chosen carefully.  For the
low-dimensional systems considered in the following, step sizes on the
order of $10^{-6}$ generally gave good results, but the optimal value is
usually not known a-priori and thus needs to be determined by trial and
error.

To avoid this problem, analytical derivative formulae can be derived.
For typical gate optimisation problems, the error function $\E(\vec{f})$
is a simple functional of the time evolution operator $X(t)$ of the
system, and a derivation similar to~\cite{25} shows that for piecewise
constant controls
\begin{equation}
  \label{eq:grad}
  \frac{\partial X(\tf)}{\partial f_{mp}}
 = X(\tf,t_p) 
   \left[\int_{t_{p-1}}^{t_{p}} X(t_p,\tau) A_m X(\tau,t_{p-1}) d\tau \right] 
   X(t_{p-1}),
\end{equation}
where $X(t,t_{p-1})=\exp[(t-t_{p-1}) \L(f_{mp})] X(t_{p-1})$ and
$\L(f_{mp})=\L_0 + \L_D + \sum_{m} f_{mp} \L_m$ for general Markovian
systems; in addition, $X(t,t_{p-1})=\exp[-i(t-t_{p-1})H(f_{mp})]
X(t_{p-1})$ and $H(f_{mp})=H_0 + \sum_{m} f_{mp} H_m$ for closed systems
subject to unitary evolution.  The latter case includes the
non-Markovian systems discussed above if $H$ is taken to be the
Hamiltonian and $X(t)$ the unitary evolution operator of the total
system including the noise qubits.  The integral in (\ref{eq:grad}) can
be evaluated exactly via the augmented matrix exponential formula
\begin{equation}
  \exp \begin{pmatrix} A & B \\ 0 & C \end{pmatrix}
 = \begin{pmatrix} 
    e^{A} & \int_0^1 e^{A(1-s)} B e^{Cs} \, ds \\ 0 & e^C
   \end{pmatrix}
\end{equation}
by setting $A=C=\L(f_{mp})\Delta t$ and $B=\L_m \Delta t$ or $A=C=-i
H(f_{mp})\Delta t$ and $B=H_m\Delta t$.  Thus we can compute both the
matrix exponential $X(t_p,t_{p-1})$ and the desired derivative simply by
computing the matrix exponential of an augmented matrix twice the size
of the system operators.  If $\L(f_{mp})$ is diagonalisable, then the
matrix exponential and gradient can alternatively be computed using the
spectral decomposition of $\L(f_{mn})$ as discussed in~\cite{25};
spectral decomposition is usually the preferred approach for such
systems.

For any Hamiltonian system, including closed systems and non-Markovian
systems with noise qubits, the total Hamiltonian $-iH(f_{mp})$ is always
skew-Hermitian and thus diagonalisable.  The superoperators for most
systems subject to Markovian non-unitary evolution, on the other hand,
are often not diagonalisable.  Therefore, the spectral decomposition
approach is usually the preferred choice for closed and non-Markovian
systems while the augmented matrix exponential is useful for other cases
where accurate gradients are required.  For small systems such as the
one- and two-qubit Markovian systems considered below, the evaluation of
the augmented matrix exponential was actually faster than using the
finite difference approximation in our simulations.  For higher
dimensional systems, the evaluation of the augmented matrix exponential
tends to be computationally more expensive than the computation of the
gradient using the finite difference approximation.  On the other hand,
the analytic formula allows for more accurate gradient computations.
Gradient accuracy is a significant factor for the performance of any
type of algorithm using gradients, especially quasi-Newton methods, and
previous work for closed systems has shown that low-order gradient
approximations, even if they are faster to compute, are detrimental to
the performance of the algorithm in terms of leading to poor convergence
and lower fidelities, unless very small step sizes are used~\cite{26,29}.
Nonetheless, with carefully chosen finite difference step-sizes it was
possible to achieve sufficient gradient accuracy to reproduce the
convergence behaviour and final fidelities obtained using the augmented
matrix exponential routine in most of our Markovian system simulations.
For non-Markovian systems, the finite difference approximation was not
used due to lack of any discernible benefit.

The algorithm also requires initial fields $\vec{f}^{(0)}$ to start the
iteration.  Here we choose the components of $\vec{f}^{(0)}$ by randomly
sampling according to a Gaussian distribution with mean $0$ and standard
deviation $\sigma(f_m^{(0)})=\delta$ where $\delta$ is a free parameter.
$\norm{\vec{f}_m}^2 = \sum_{p=1}^K |f_{mp}|^2 \approx K\delta^2$ shows
that $\delta$ determines the norm of the initial field vector of each
component field $\vec{f}_m^{(0)}=(f_{m1},\ldots,f_{mK})$.  For piecewise
constant functions with fixed $\Delta t$, the norm of the field vector is
proportional to the $L^2$ norm of the field as a function over $[0,\tf]$:
\begin{equation}
  \norm{\vec{f}_m(t)}_2^2
  = \int_0^\tf |f_m(t)|^2 dt 
  = \Delta t \sum_{p=1}^K |f_{mp}|^2 = \Delta t \norm{\vec{f}_m}^2.
\end{equation}
We refer to the square of the $L^2$ norm of a field as the fluence of
the pulse, a quantity that is of independent physical interest as it is
proportional to the pulse energy.

\subsection{Explicit Error Functionals for the Markovian/Non-Markovian Case}

For general Markovian dynamics, the most natural way to compare quantum
processes is by considering the distance between the processes in the
adjoint representation.  The adjoint representation of a unitary target
operator $W$ with respect to the chosen basis $\{\sigma_k\}$ is given by
the $N_1^2\times N_1^2$ real matrix $Y$ with $Y_{mn}= \Tr(\sigma_m W
\sigma_n W^\dag)$.  Letting $X(t)$ be the real adjoint representations
of the time-evolution operator of the system defined earlier, we can
define the gate error in terms of the square of the Frobenius (or
Hilbert-Schmidt) norm of $\Lambda(t)=Y-X(t)$,
\begin{equation}
 \E_1'  = \lambda \Tr[\Lambda^\tp(\tf)\Lambda(\tf)]
        = \lambda \left(\Tr(Y^\tp Y) + \Tr[X(\tf)^\tp X(\tf)]-2\Tr[Y^\tp X(\tf)] \right),
\end{equation}
where $\lambda$ is a scaling factor.  The gate error is a simple
functional of the evolution operator and thus
\begin{equation}
 \frac{\partial \E_1'}{\partial f_{mp}}
 = 2 \lambda \Tr\left[\Lambda^T(T) \frac{\partial \Lambda(T)}{\partial f_{mp}}\right]
 = -2 \lambda \Tr\left[\Lambda^T(T) \frac{\partial X(T)}{\partial f_{mp}}\right],
\end{equation}
where $\tfrac{\partial X(T)}{\partial f_{mp}}$ are the partial
derivatives of the evolution operator as defined in (\ref{eq:grad}).  We
will use this performance index with $\lambda=\tfrac{1}{2N_1^2}$ for the
optimisation.  However, for the sake of a direct error comparison with 
closed and non-Markovian systems, we also define the error functional
\begin{equation}
  \E_1 = 1- \F_1, \quad \F_1 = \sqrt{1-\E_1'},
\end{equation}
where $\F_1$ is the \emph{unit gate fidelity}.  It is easy to check that 
this gate fidelity agrees with the gate fidelity for closed systems if
the Lindblad operators vanish.

For non-Markovian systems, one may be tempted to simply define the gate
error by
\begin{equation}
  \label{eq:F_closed}
  \E = 1 - \tfrac{1}{N} |\Tr[ (W \otimes \ONE)^\dag X(\tf)]| ,
\end{equation}
where $N$ is the dimension, $X(\tf)$ the (unitary) time evolution
operator of the composite system, $W$ the target operator on the system,
and $\ONE$ the identity operator on the noise subsystem.  However, this
error vanishes only if we simultaneously implement the target gate $W$
on the system subsystem and the identity on the noise subsystem.
Usually, however, we do not care what the evolution of the noise
subsystem is.  Therefore, a better measure of the gate error is
\begin{equation}
  \label{eq:E2}
  \E_2 = \lambda \min_{\phi} \norm{W \otimes \phi - X(\tf)}^2,
\end{equation}
where $\phi$ is an arbitrary time-evolution operator acting on the noise
subsystem only, and $\lambda$ is a normalisation constant that can be
chosen so that the error ranges between $0$ and $1$, for instance.%
\footnote{The distance measure (\ref{eq:E2}) agrees with the one
introduced by Grace et al. in~\cite{6} except that the right-hand side
is squared to make the non-Markovian gate fidelity $1-\E_2$ agree with
the usual gate fidelity for closed systems in the absence of noise qubits.}
Taking the matrix norm in~(\ref{eq:E2}) to be the Frobenius norm and
setting $\lambda=\frac{1}{2N}$, it can be shown that~\cite{6,27}
\begin{equation}
  \label{eq:E2b}
  \E_2 = 1- \F_2, \quad \F_2 = \tfrac{1}{N} \Tr \sqrt{Q^\dag Q}, 
\end{equation}
where $Q = \Tr_S( (W\otimes \ONE)^\dag X(\tf))$, $\Tr_S$ denotes the
partial trace over the system $S$, and $\F_2$ is the gate fidelity.
Differentiating (\ref{eq:E2b}) with respect to $f_{mp}$, we obtain
\begin{align}
  \frac{\partial \E_2}{\partial f_{mp}}
  = -\frac{1}{N} \Re\Tr \left( \frac{1}{\sqrt{Q^\dag Q}} Q^\dag
                               \frac{\partial Q}{\partial f_{mp}}\right),\ \quad
  \frac{\partial Q}{\partial f_{mp}}
  = \Tr_S \left[ (W \otimes \ONE)^\dag \frac{\partial X(\tf)}{\partial f_{mp}} \right],
\end{align}
and thus the gradient is again defined in terms of the partial
derivatives of the total evolution operator $X(\tf)$ with regard to the
control variables $f_{mp}$.  $\sqrt{Q^\dag Q}$ may become singular but
from the singular value decomposition $Q=G D E^\dag$, where $G$ and $D$
are unitary matrices and $D$ is a diagonal matrix with non-negative
entries, one can easily verify that $(Q^\dag Q)^{-1/2} Q^\dag = E
G^\dag$, which can be defined even if $\sqrt{Q^\dag Q}$ is singular.

\section{Results and Discussion}

To evaluate the performance of the algorithm and quality of the
solutions found, we studied a variety of gate optimisation tasks for
single, two-, and three-qubit systems in different Markovian and
non-Markovian environments.  Details of the system parameters chosen are
given in Tables \ref{table1} and \ref{table2}.  In our system of units,
$\hbar=\omega_1 =|\mu_1| = k_B = 1$, where $\omega_1$ is the angular
frequency of the system qubit in the one-qubit case and $k_B$ is the
Boltzmann constant. Throughout this paper time is expressed in units of
$\omega_1^{-1}$.  In addition to fundamental issues such as asymptotic
error values, how rapidly we converge to these, and the dependence of
the convergence behaviour and limiting values on algorithmic parameters
such as the choice of the initial fields $\vec{f}^{(0)}$, we also
considered the impact of unwanted side effects such as field leakage and
analysed the characteristics of the optimal fields.

\begin{table*}
\caption{Summary of the system parameters in the Markovian model}
\label{table1}
\begin{tabular}{|p{1in}|p{1.1in}|p{1.1in}|p{1.1in}|}
\hline
& \textbf{1 system qubit} & \textbf{2 system qubits} & \textbf{3 system qubits}
\\\hline
\textbf{Decoherence type} (emission / dephasing rate in units of $\omega_1$)
&
\begin{minipage}[t]{1.2in}
\raggedright
No decoherence 
Spontaneous emission (0.02)\\
Independent dephasing in the z-basis (0.02)\\
Independent dephasing in the x-basis (0.02)
\end{minipage}
&
\begin{minipage}[t]{1.2in}
\raggedright
No decoherence\\
Spontaneous emission (0.02)\\
Independent dephasing in the z-basis (0.02)\\
Independent dephasing in the x-basis (0.02)
\end{minipage}
&
\begin{minipage}[t]{1.2in}
\raggedright
No decoherence\\
Independent dephasing in the z-basis (0.02)\\
Independent dephasing in the x-basis (0.02)\\
Correlated dephasing in the z-basis (0.02)
\end{minipage}
\\\hline
\textbf{Frequencies of system qubits} (in units of $\omega_1$)
& 1 & 0.95 and 1.05 & 0.95, 1, and 1.05
\\\hline
\textbf{Heisenberg coupling strength} (in units of $\omega_1$)
& 1 & 1 & 1 
\\\hline
\textbf{Target gate(s)}
& Hadamard, Identity, T-gate & Identity, CNOT &
3-qubit quantum Fourier transform (QFT), Identity
\\\hline
\textbf{Target times} in units of $\omega_1^{-1}$, (Time slices $K$)
& 5 (25), 25 (25)
& 25 (150), 50 (150), 75 (150), 100 (150)
& 150 (300)
\\\hline
\textbf{Initial field} std deviation
& 0.1, 1, or 10
& 0.01, 0.1, 1, or 10
& 1
\\\hline
\end{tabular}
\end{table*}

\begin{table*}
\caption{Summary of the non-Markovian systems analysed}
\label{table2}
\begin{tabular}{|p{0.9in}|p{1.2in}|p{1.2in}|p{1.2in}|}
\hline
& \textbf{1 system qubit} & \textbf{2 system qubits} & \textbf{3 system qubits} 
\\\hline
\textbf{Number of noise qubits}
& 0, 1, 2, 4, or 6
& 0, 1, 2, or 4
& 0 or 2 
\\\hline
\textbf{Frequencies} for system qubits (in units of $\omega_1$)
& 1
& 0.95 and 1.05
& 0.95, 1, and 1.05
\\\hline
\textbf{Noise qubit frequencies} (in units of $\omega_1$)
& 
\begin{minipage}[t]{1.2in}
$(\pi-2.14)^{-1}$, $(\pi-2.14)$
$(\pi-2.1)^{-1}$, $(\pi-2.1)$
$(\pi-2)^{-1}$, $(\pi-2)$
\end{minipage}
& 
\begin{minipage}[t]{1.2in}
$(\pi-2.14)^{-1}$, $(\pi-2.14)$\\
$(\pi-2.1)^{-1}$, $(\pi-2.1)$ 
\end{minipage}
&
$(\pi-2.14)^{-1}$, $(\pi-2.14)$
\\\hline
\textbf{Heisenberg coupling strengh $\gamma$} (in units of $\omega_1$)
& 0.02 between system and noise qubits
& 0.1 between system qubits, 0.01 between system and noise qubits
& 0.1 between system qubits, 0.01 between system and noise qubits
\\\hline
\textbf{Target gates}
& Hadamard, Identity, T-gate
& Identity, CNOT
& 3-qubit quantum Fourier transform, Identity
\\\hline
\textbf{Target times} in units of $\omega_1^{-1}$, \textbf{(time slices)}
& 2 (25), 3 (25), 4 (25), or 25 (25)
& 25, 30, 35, 40, 45, 50, 55, 60, 65, 70, 75, 80, 85, 90, 95, 100, 125 
  (150 time slices each)
& 150 (300) or 300 (300)
\\\hline
\textbf{Initial field} std deviation
& 0.1, 1, or 10
& 0.01, 0.05, 0.1, 0.5, 1, 5, or 10
& 1
\\\hline
\end{tabular}
\end{table*}

\subsection{Convergence Behaviour and Asymptotic Error Values}

\begin{figure*}
\caption{Convergence behaviour}\label{fig:conv}
\subfloat[\sf Two-qubit CNOT gate, no decoherence, 
$\tf=75$~$\omega_1^{-1}$,~$\sigma(f^{(0)})=0.1$]{
\includegraphics[width=0.49\textwidth]{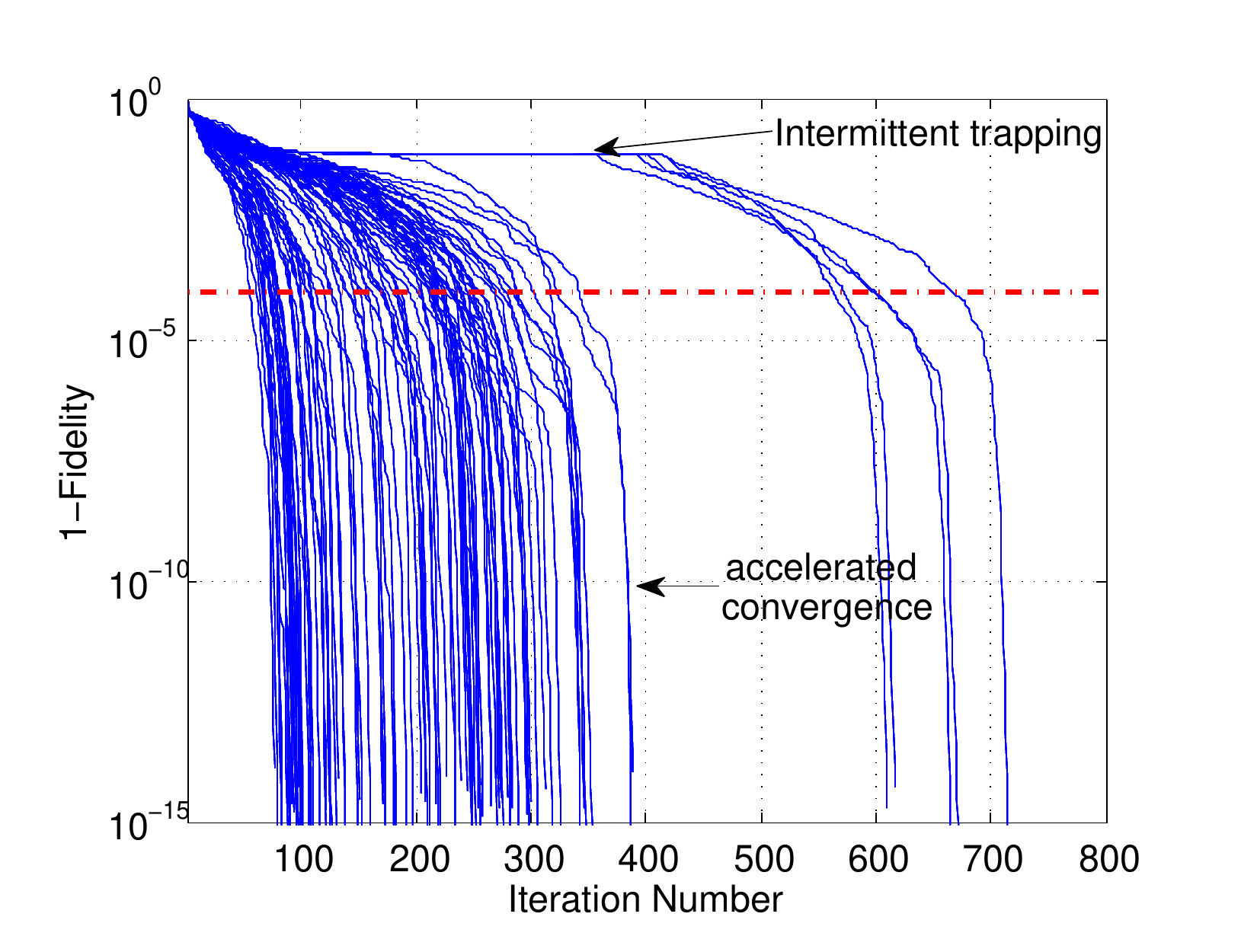}}
\subfloat[\sf One-qubit T gate, spontaneous emission,  
$\tf=5$~$\omega_1^{-1}$,~$\sigma(f^{(0)})=0.1$]{
\includegraphics[width=0.49\textwidth]{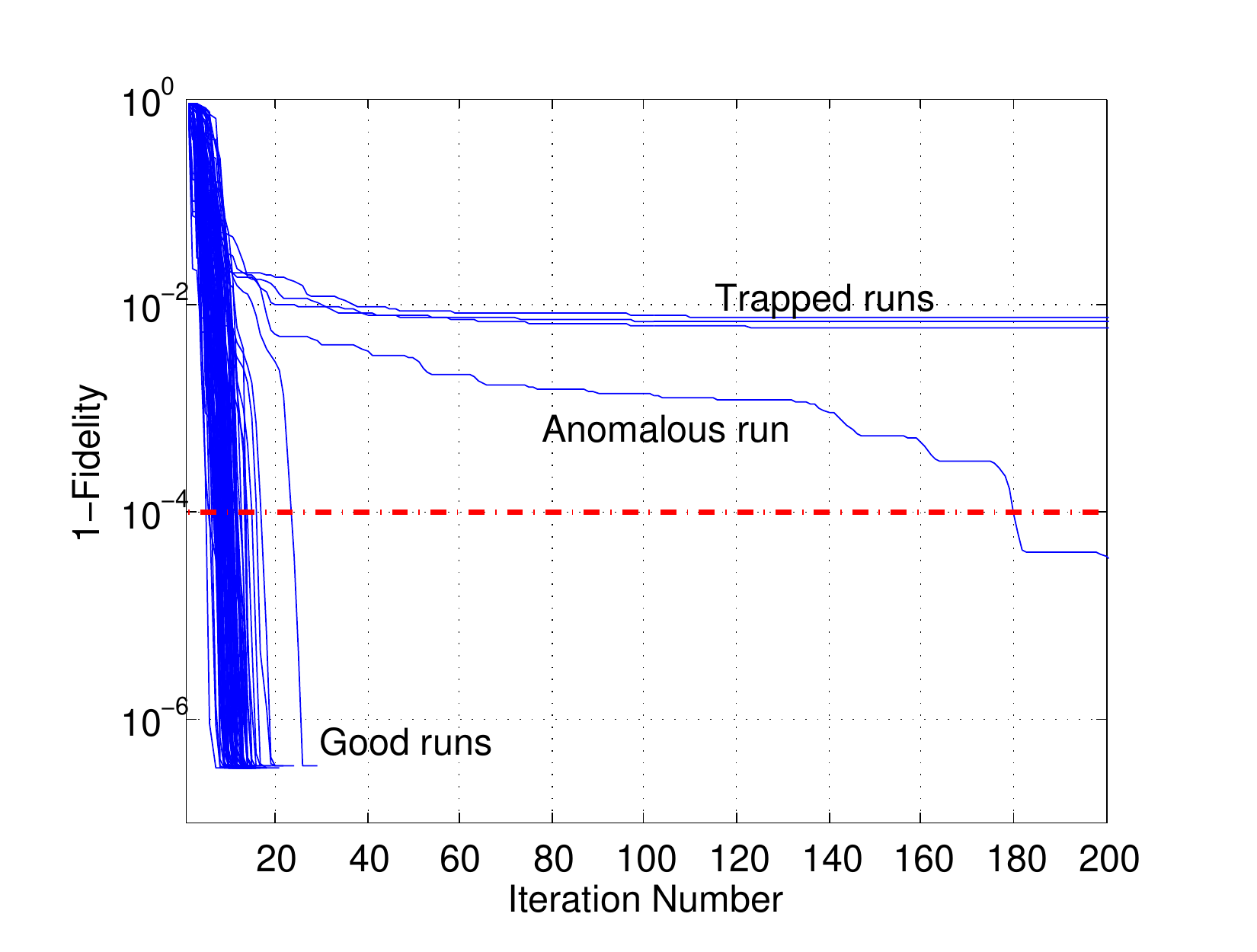}}
\\[3ex]
\subfloat[\sf One-qubit Had gate, $x$-dephasing,
 $\tf=5$~$\omega_1^{-1}$,~$\sigma(f^{(0)})=10$]{
\includegraphics[width=0.49\textwidth]{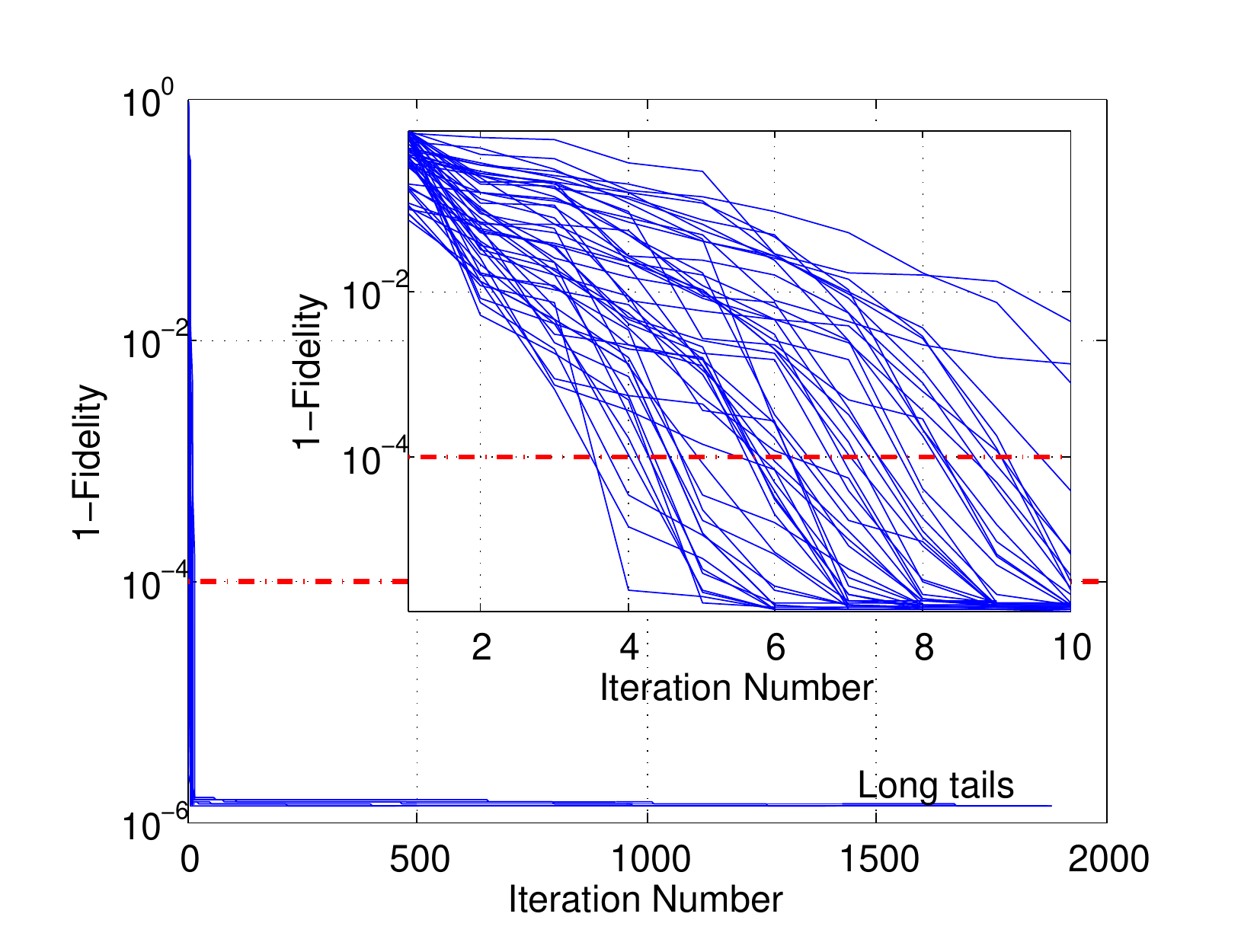}} 
\subfloat[\sf One-qubit Had gate, two noise qubits, 
$\tf=3$~$\omega_1^{-1}$, $\norm{\vec{f}^{(0)}}=10$]{
\includegraphics[width=0.49\textwidth]{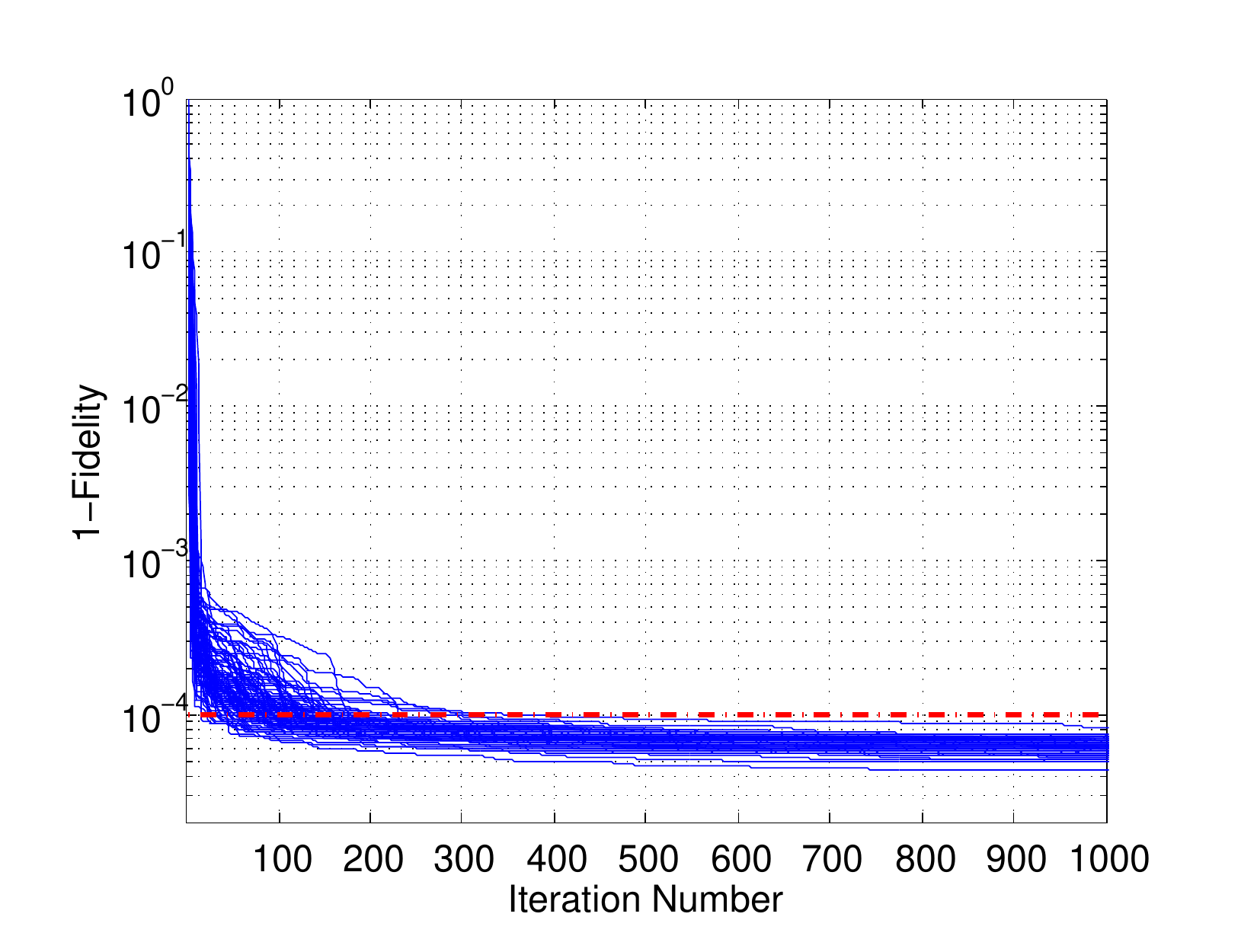}}
\\[3ex]
\subfloat[\sf Three-qubit QFT gate, spontaneous emission,
$\tf=150$~$\omega_1^{-1}$,~$\sigma(f^{(0)})=1$]{
\includegraphics[width=0.49\textwidth]{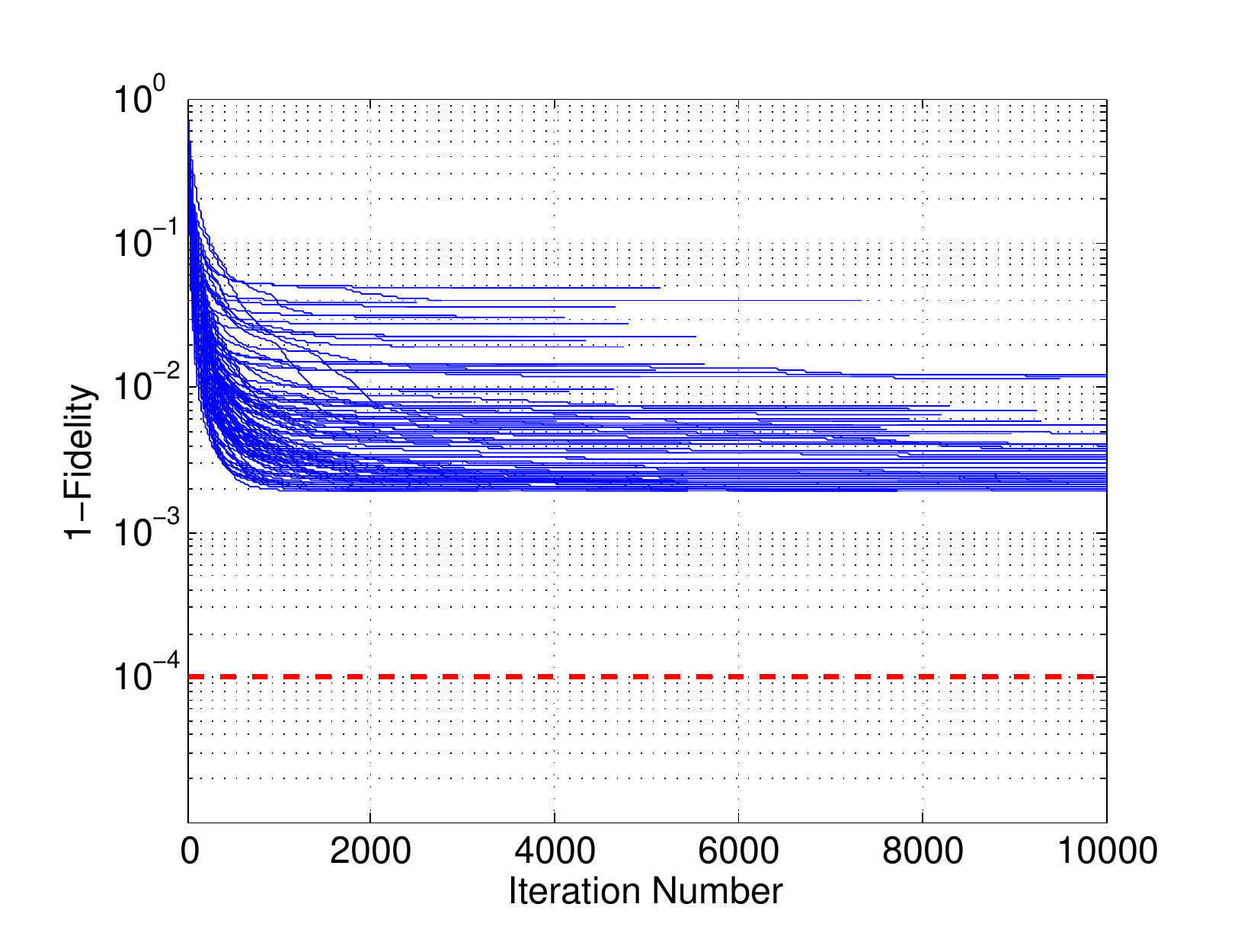}}
\subfloat[\sf Three-qubit QFT gate, two noise qubits, 
$\tf=300$~$\omega_1^{-1}$, $\norm{\vec{f}^{(0)}}=1$]{
\includegraphics[width=0.49\textwidth]{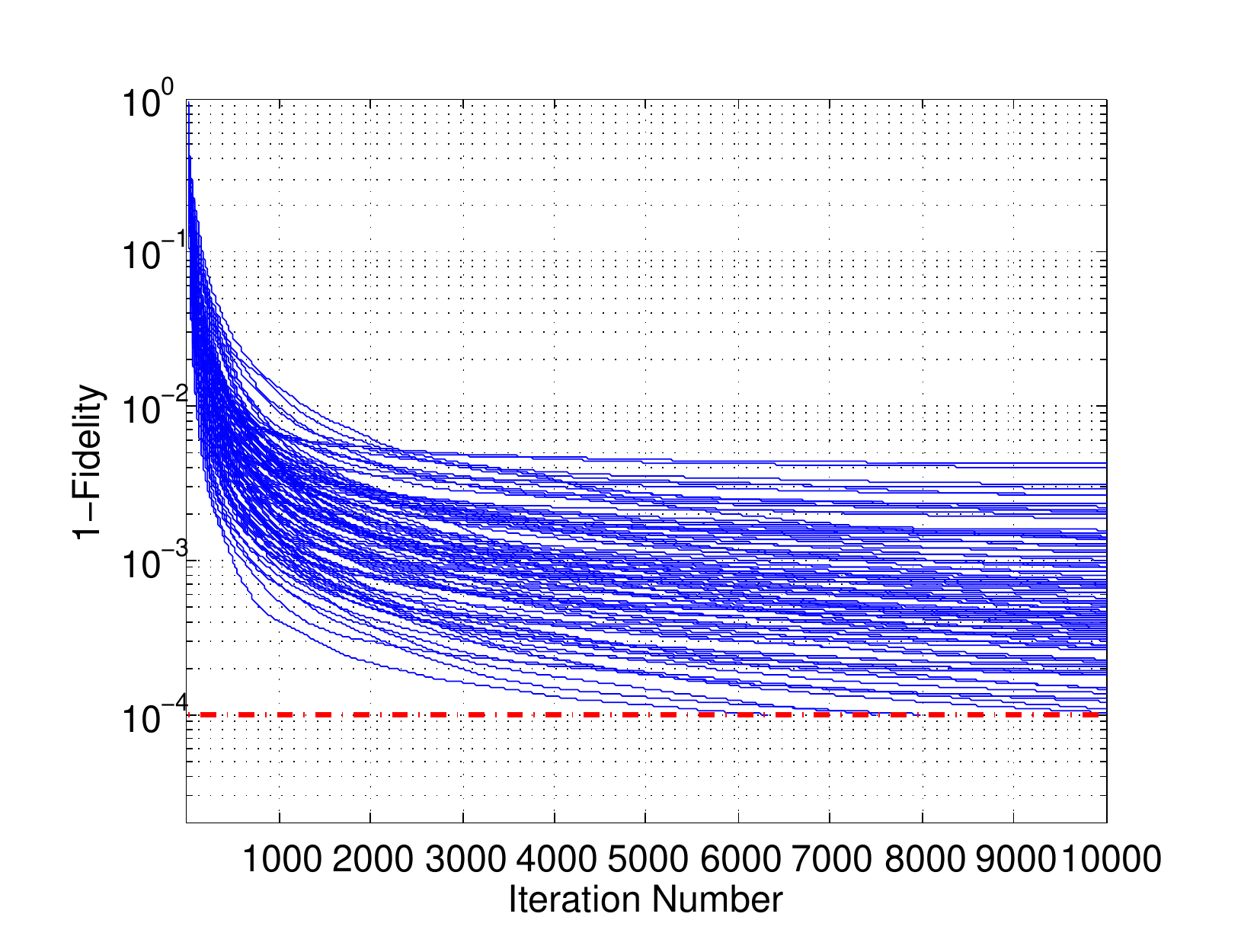}}
\end{figure*}

For simple problems, we found considerable uniformity in terms of the
final values of the fidelities achieved for different runs and different
target gates.  For example, for the one- and two-qubit gate problems
considered the final fidelities of more than 95\% of all runs for a
given problem typically agreed to three or more decimal places.  This
suggests that most runs converged to a control that achieved close to
optimal performance in terms of reaching an error very close to the
global minimum attainable for the given form of decoherence and control
field restrictions (such as the field parameterisation, target time, and
time resolution).  This was the case in both the Markovian and
non-Markovian case although the errors in the Markovian case were
higher, especially for the more complex gates.  For instance, for the
three-qubit QFT gate the maximum asymptotic fidelity was around 99.8\%
for spontaneous emission, 99.3\% for dephasing in either the $x$ or $z$
basis, and 99.2\% for correlated $zz$-dephasing at a rate of $0.02$ for
$T=150$, while errors $<10^{-4}$ were attainable for the non-Markovian
cases.  This is not unexpected considering that the Markovian setting
lacks the possibility of coherence revivals.  As the difficulty of the
gate optimisation problems increased, however, far less uniform
convergence behaviour and increasingly large spreads of final fidelities
were observed, again in both the Markovian and non-Markovian cases.
This is exemplified in Fig.~\ref{fig:conv} which shows the convergence
behaviour for 100 runs each for different cases.  For instance, the
error of the best run in Fig.~\ref{fig:conv}(f) was $<10^{-4}$ while the
error of the worst run was $0.0043$, two orders of magnitude larger.

\textit{Conclusion 1:} While for simple control tasks a single run with
a randomly chosen initial field usually suffices, multiple runs with
different initial conditions are essential to find a control that
achieves the best possible fidelity for harder control problems.

An important difference in the convergence behaviour between open and
closed systems is that in the latter we generally observed accelerated
convergence as shown in Fig.~\ref{fig:conv}(a), while in the dissipative
case the convergence was at best linear as in Fig.~\ref{fig:conv}(b)
and, in most cases, we actually observed a slowdown in the rate at which
the error decreased.  Moreover, trapping and long tails are seen in the
Markovian and non-Markovian cases.  This behaviour is consistent with
convergence to a limiting value of the error strictly greater than $0$,
which is expected for open systems.

The accuracy of the gradient approximation is another limiting factor
for very high fidelities. When quasi-Newton methods are used as in
our case, the errors in the gradients will eventually also lead to large
errors in the approximate Hessian and increasingly poor performance of
the quasi-Newton iteration, precluding further reductions in the errors.
The algorithm should be terminated before this regime is reached.  The
use of analytic gradient formulae can alleviate this problem.  However,
even the augmented matrix exponential or spectral decomposition gradient
formulae, though theoretically exact, will be subject to numerical
approximation errors in practice.
\footnote{As an aside, we initially suspected that long tails observed
in the Markovian case were the result of a loss of accuracy of the
finite difference gradient approximation used for the initial runs in
the Markovian case.  However, the convergence behaviour was mostly
unchanged when the analytic gradient formula was used and the tails
persisted, suggesting that the finite-difference approximation error was
not a limiting factor in these simulations.}

\textit{Conclusion 2:} Trapping, long tails, and diminishing returns as a
function of the iteration number for open systems make sensible termination
conditions \emph{essential}.

While for controllable closed systems it is often reasonable to set a
threshold such as $-10^{-4}$ for the error as the termination condition, for
open systems there are usually no strict upper bounds on the attainable
fidelities and the best strategy therefore is to \emph{dynamically
monitor} the rate of decrease in the error and terminate the optimisation when this value
becomes too low and we have reached an asymptotic regime.

\subsection{Dependence on Gate Operation Time and Initial Fields}

Assuming the model and objective have already been chosen, the algorithm depends
on two key inputs: the gate operation time $\tf$ and the choice of the
initial fields. An interesting question is how the choice of these
parameters affects the convergence behaviour and limiting values of the
fidelities attained.  Initial test runs suggested a dependence of the
convergence behaviour on the magnitude of the initial fields, which in
our case was determined by the variance of the Gaussian distribution we
sampled from.  To better understand the effects of the target time and
magnitude of the initial fields on the convergence behaviour and
fidelities attained, we combined the data from many runs for initial
fields with different norms (always sampled from a normal distribution)
and different target times into 2D density plots for the \emph{success
rate} and the \emph{success speed}.  The success rate here is the
fraction of runs for a given choice of initial field norm and target
time that reached fidelities at or above the error threshold, which we
set to be $10^{-4}$.  The success speed was defined as the inverse of
the expected time required to succeed in finding a control that achieves
the desired error threshold, where the expected time to succeed is
computed as~\cite{29}
\begin{equation}
 \mbox{mean failed run time} \times
  \frac{ \mbox{number of failed runs}}{\mbox{number of successful runs}}
   + \mbox{mean successful run time}.
\end{equation}

\begin{figure*}
\caption{Success rate plots for two-qubit system for error threshold
 $10^{-4}$ with colour bars indicating the success rate}
\label{fig:success-rate}
\subfloat[\sf Two-qubit CNOT gate (no noise qubits)]
{\includegraphics[width=0.42\textwidth]{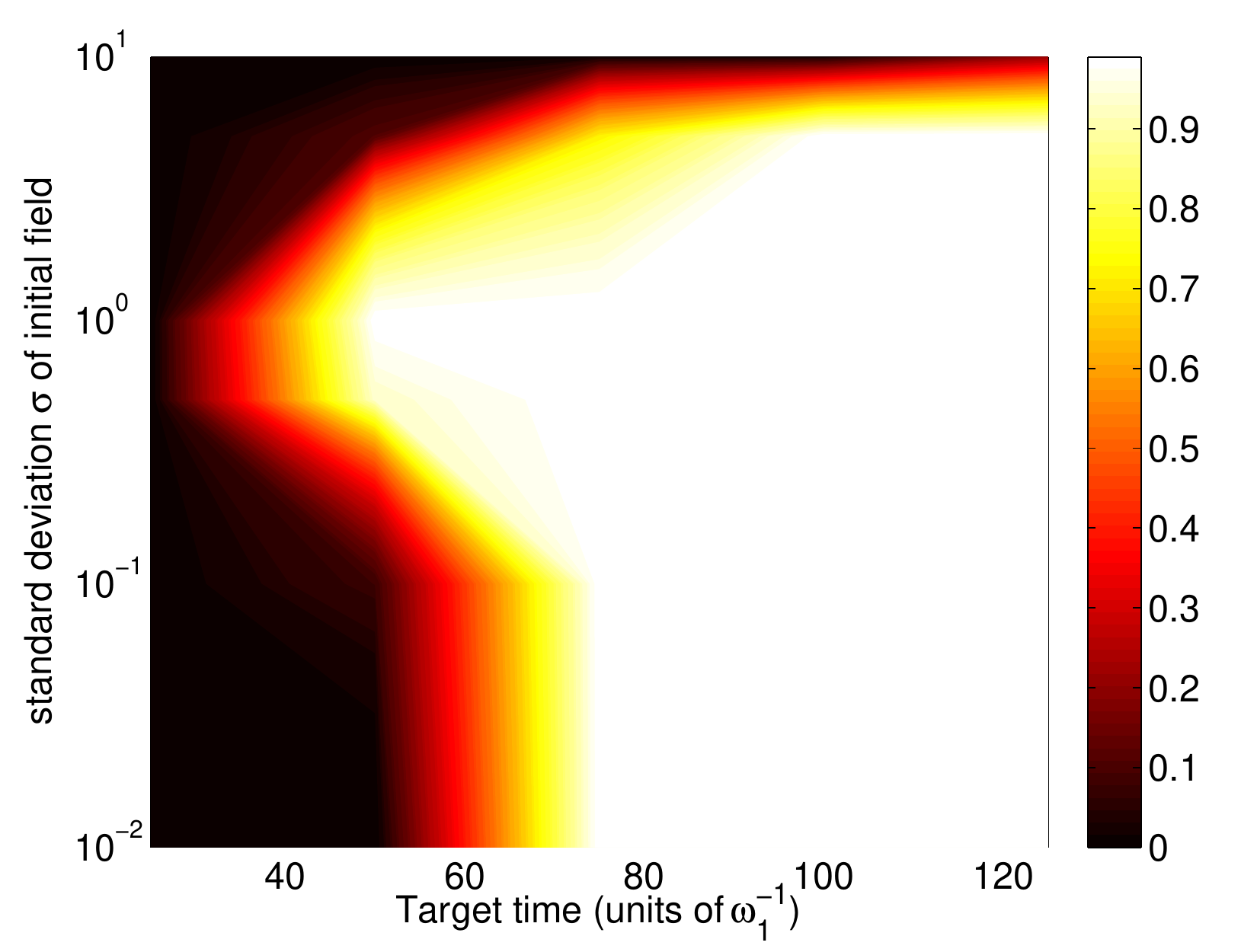}} 
\hfill
\subfloat[\sf Two-qubit Id gate (no noise qubits)]
{\includegraphics[width=0.42\textwidth]{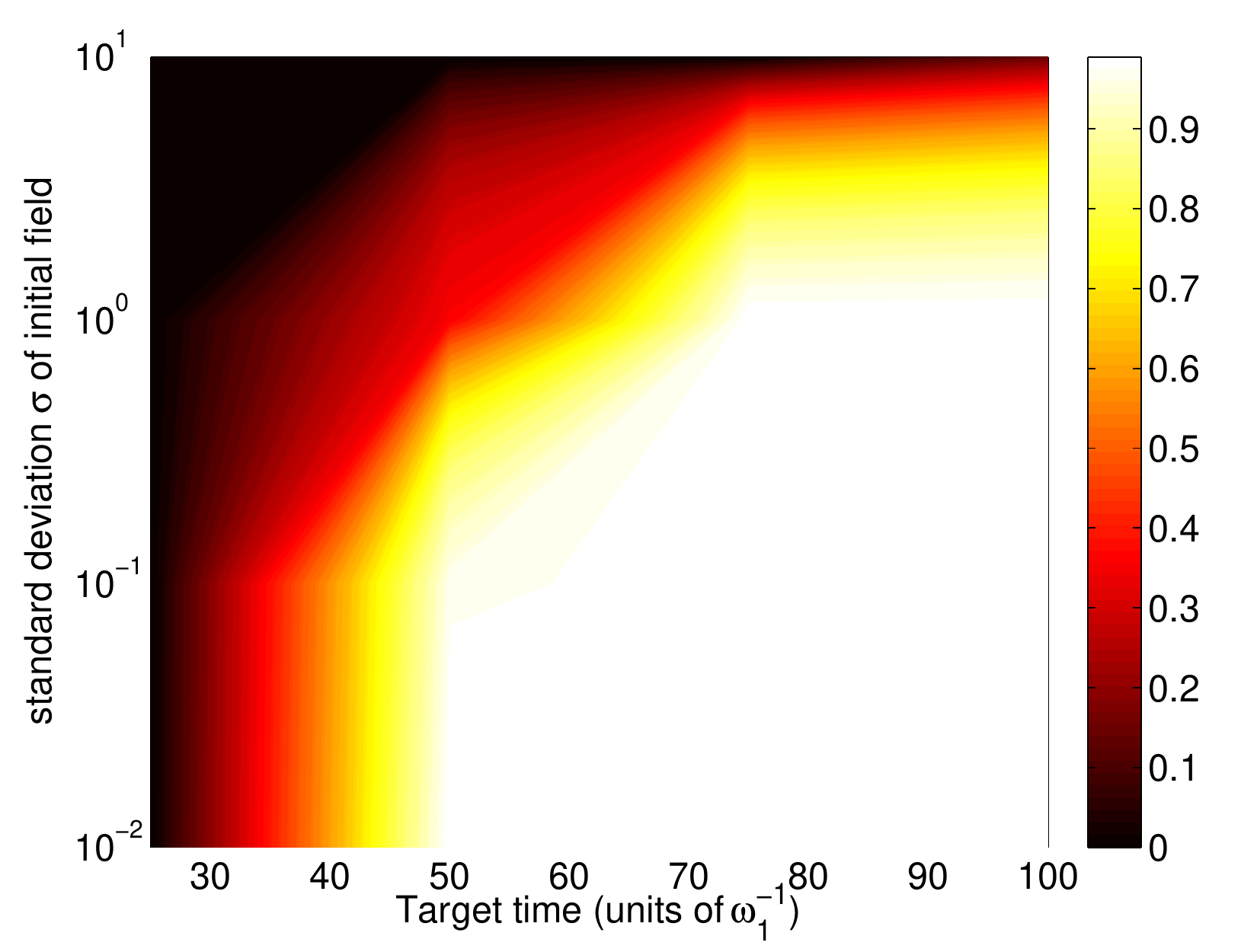}} 

\subfloat[\sf Two-qubit CNOT gate (1 noise qubit)]
{\includegraphics[width=0.42\textwidth]{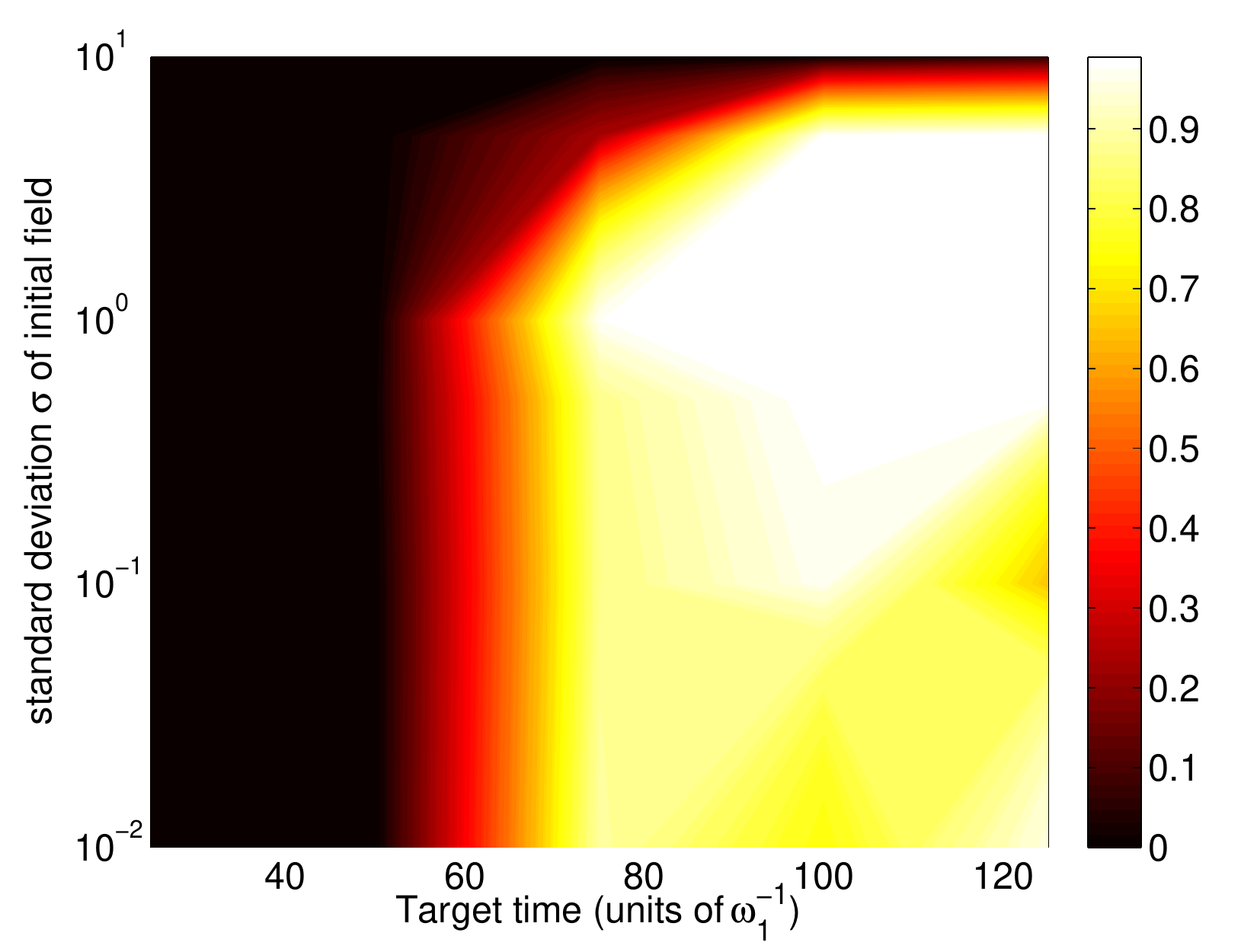}}
\hfill
\subfloat[\sf Two-qubit Id gate (1 noise qubit)]
{\includegraphics[width=0.42\textwidth]{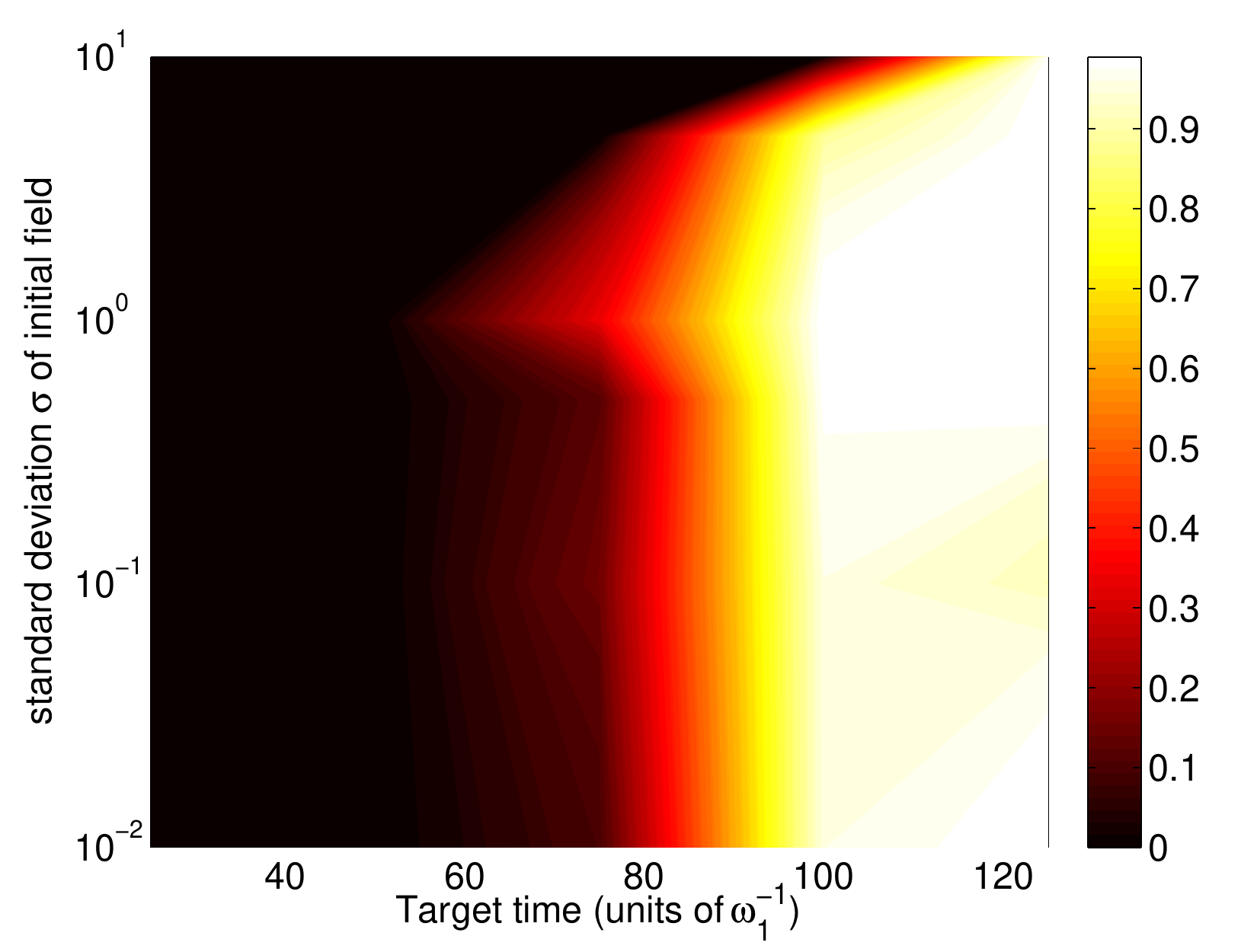}} 

\subfloat[\sf Two-qubit CNOT gate (2 noise qubits)]
{\includegraphics[width=0.42\textwidth]{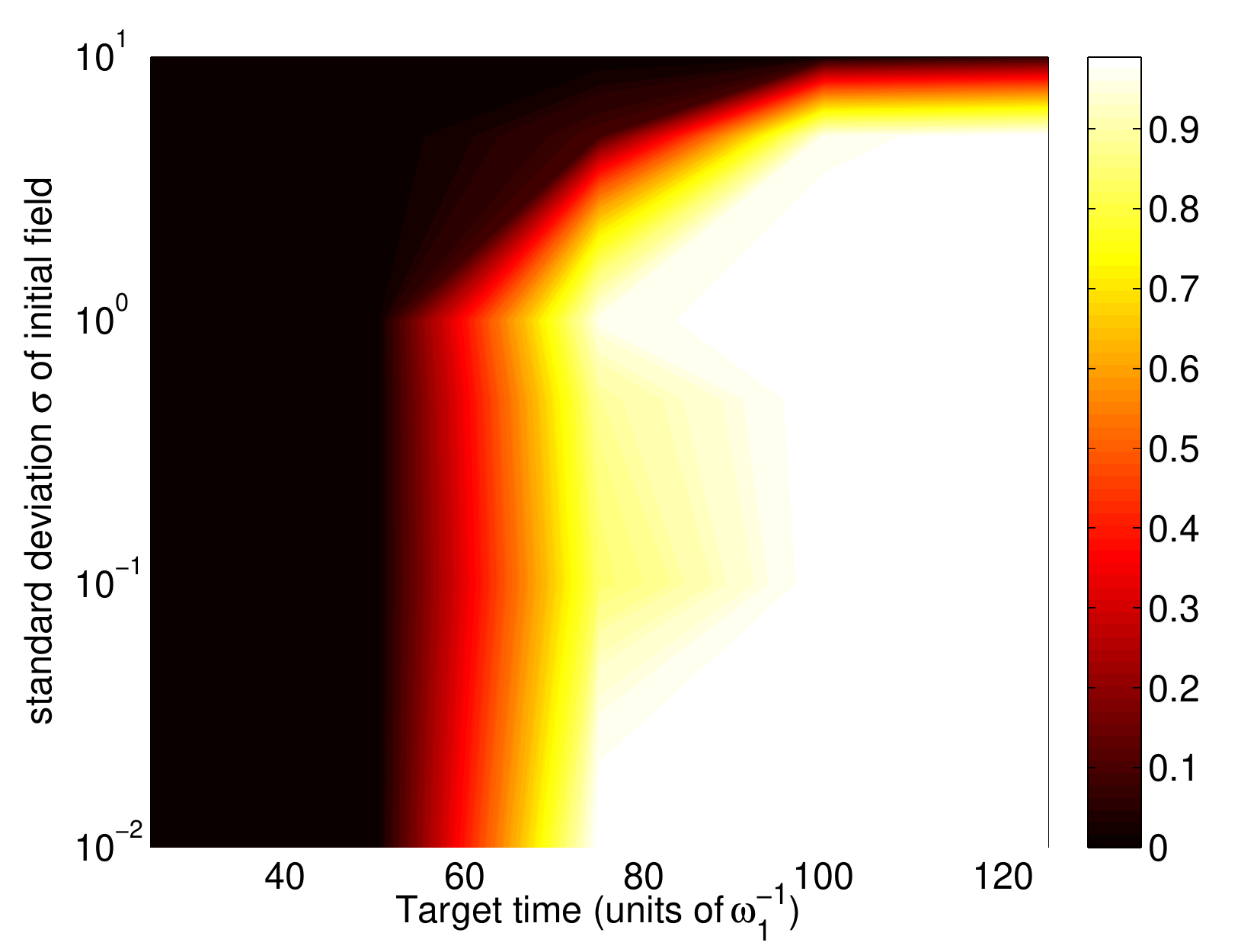}}
\hfill
\subfloat[\sf Two-qubit Id gate (2 noise qubits)]
{\includegraphics[width=0.42\textwidth]{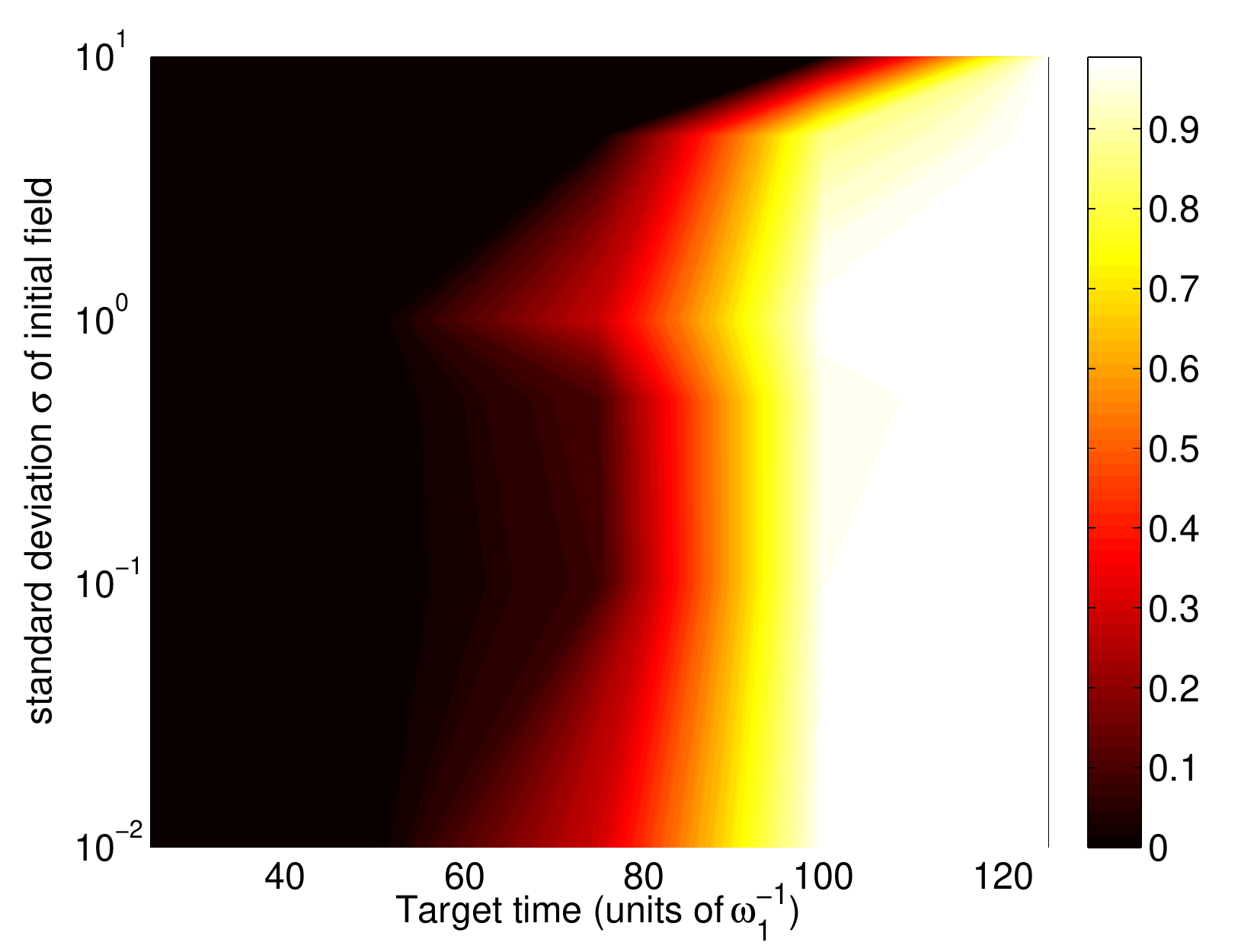}}

\subfloat[\sf Two-qubit CNOT gate (4 noise qubits)]
{\includegraphics[width=0.42\textwidth]{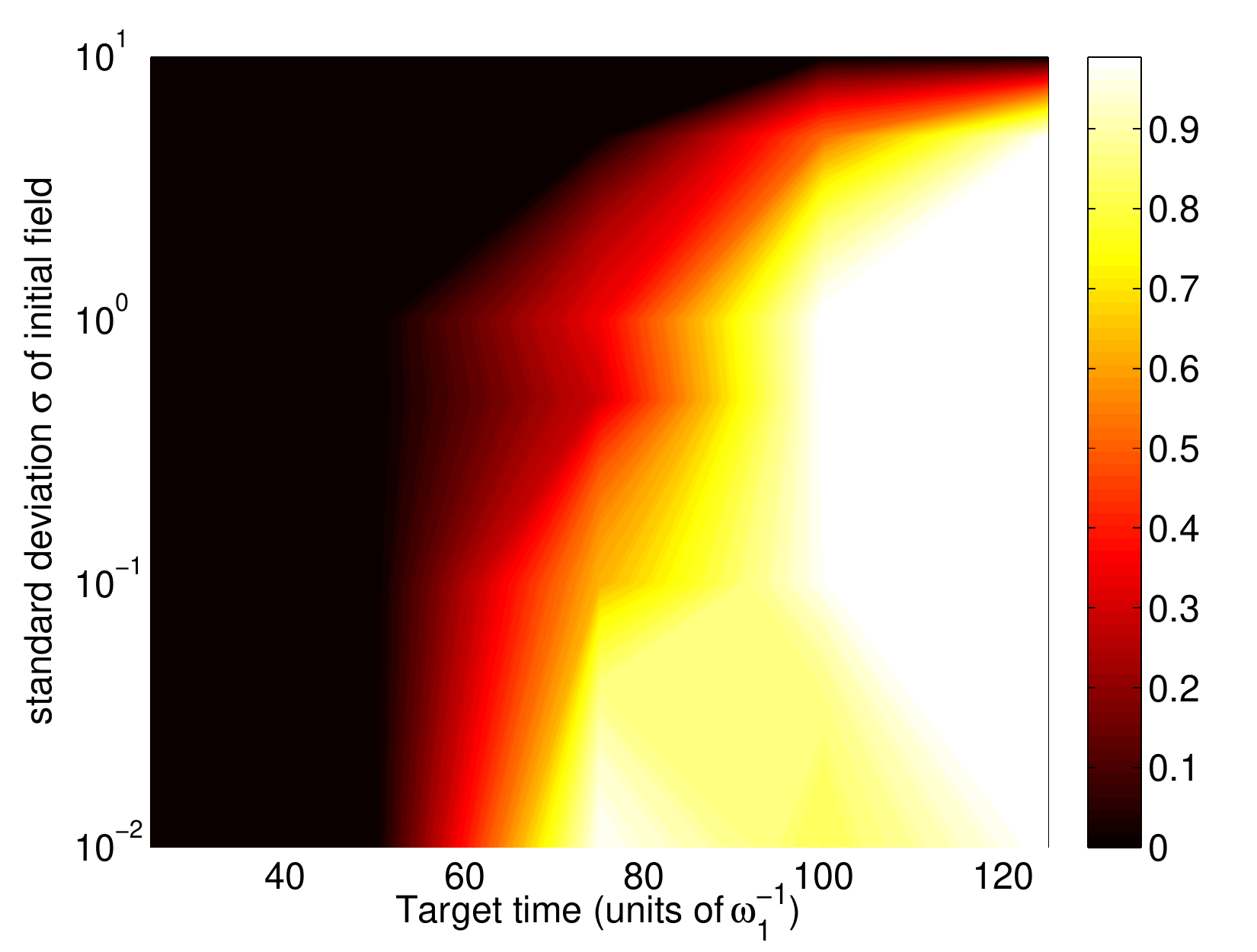}}
\hfill
\subfloat[\sf Two-qubit Id gate (4 noise qubits)]
{\includegraphics[width=0.42\textwidth]{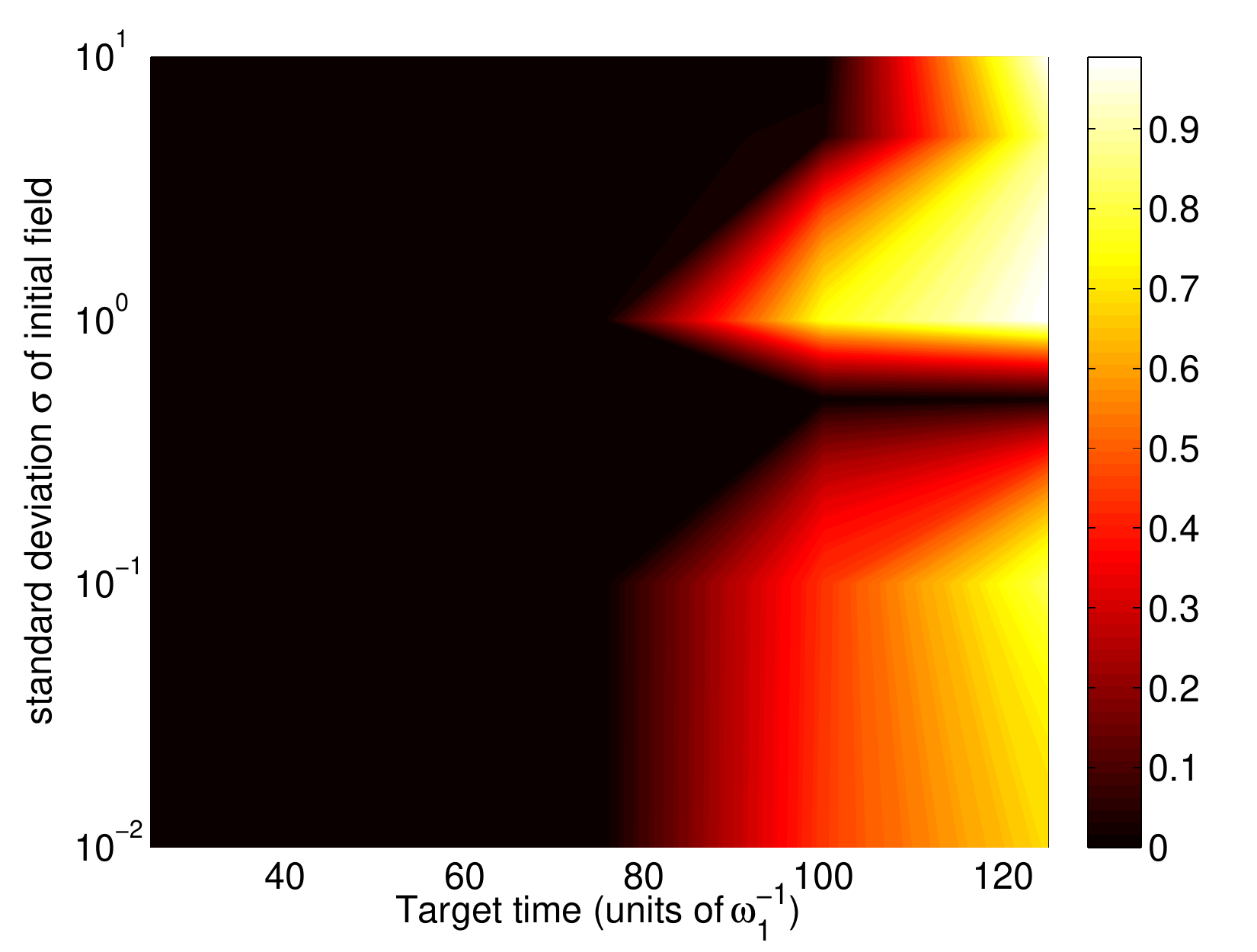}}

\subfloat[\sf CNOT gate, $z$-dephasing]
{\includegraphics[width=0.42\textwidth]{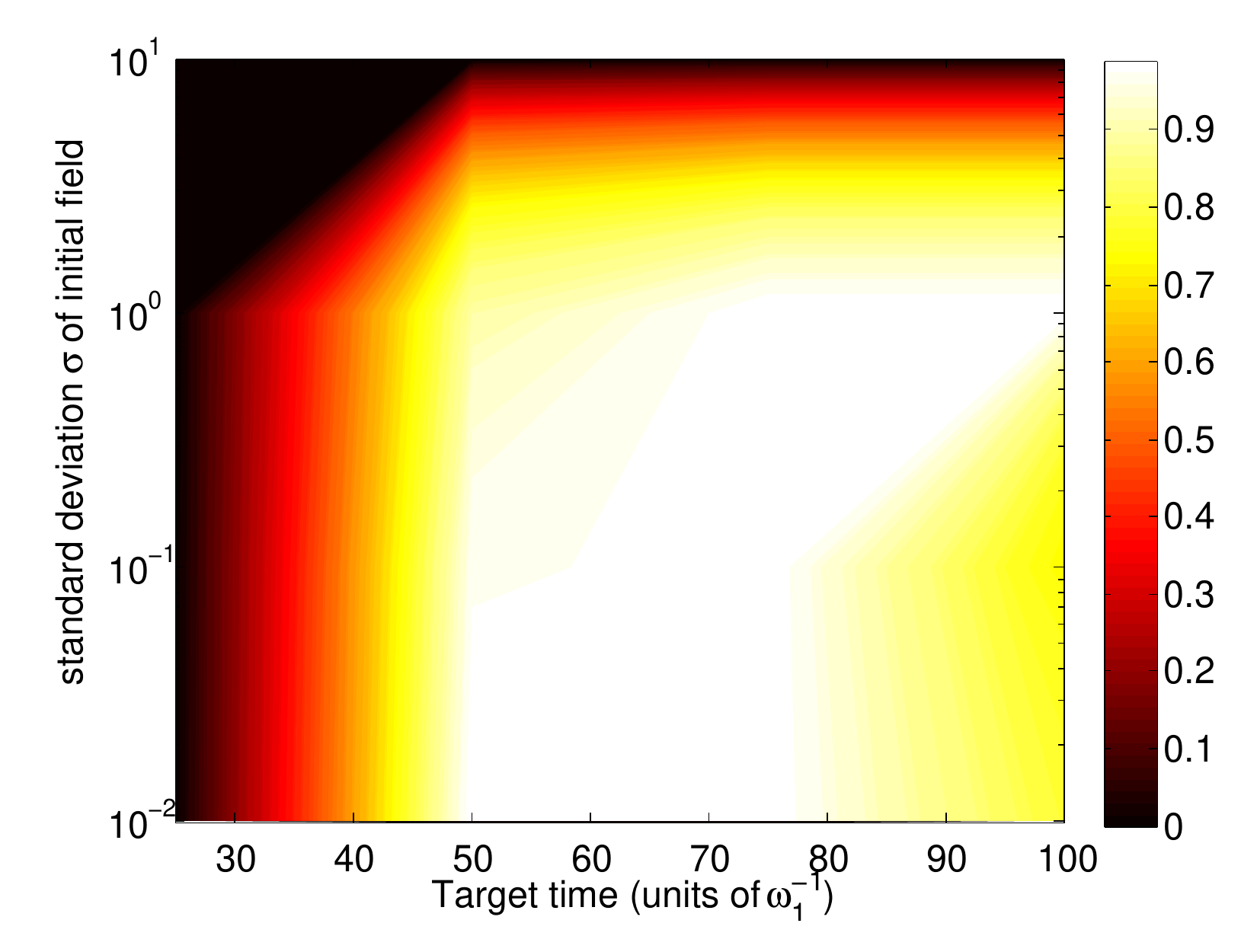}} 
\hfill
\subfloat[\sf Two-qubit Id gate, $z$-dephasing]
{\includegraphics[width=0.42\textwidth]{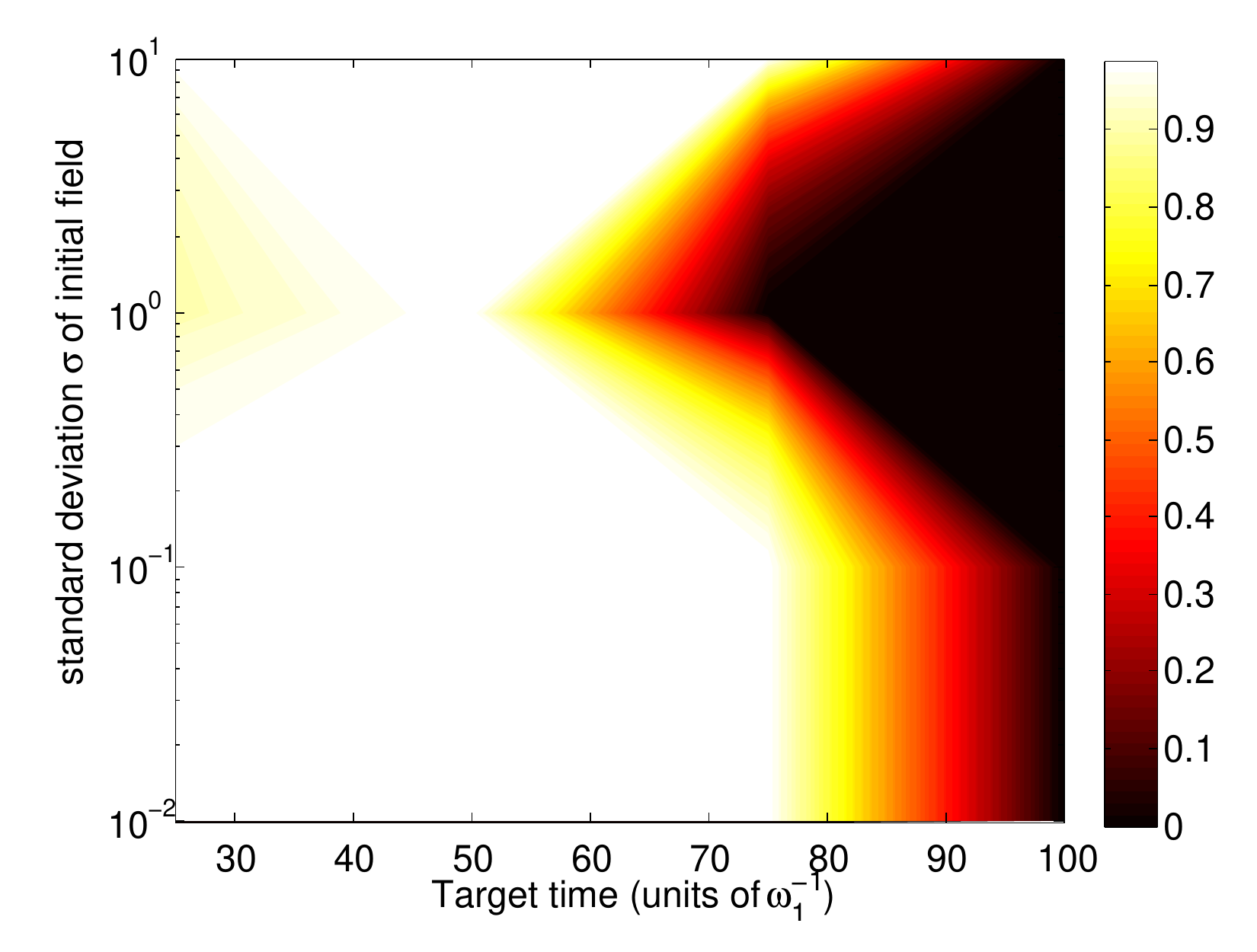}} 
\end{figure*}

\begin{figure*}
\caption{Success speed plots for two-qubit system for error threshold
$10^{-4}$ with colorbars indicating success speed.}  \label{fig:success-speed} 

\subfloat[\sf Two-qubit CNOT gate(no noise qubits)]
{\includegraphics[width=0.42\textwidth]{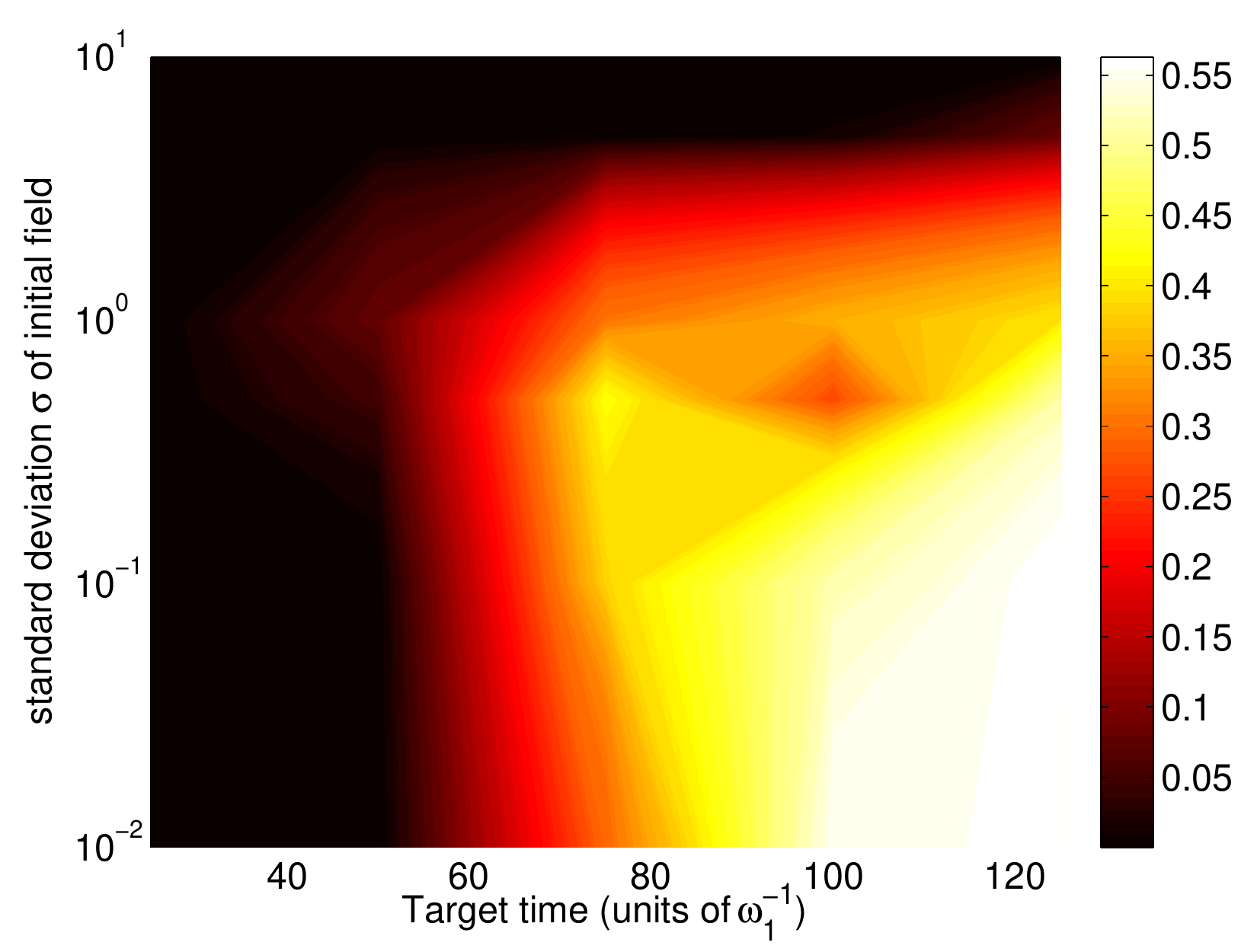}}
\hfill
\subfloat[\sf Two-qubit Id gate (no noise qubits)] 
{\includegraphics[width=0.42\textwidth]{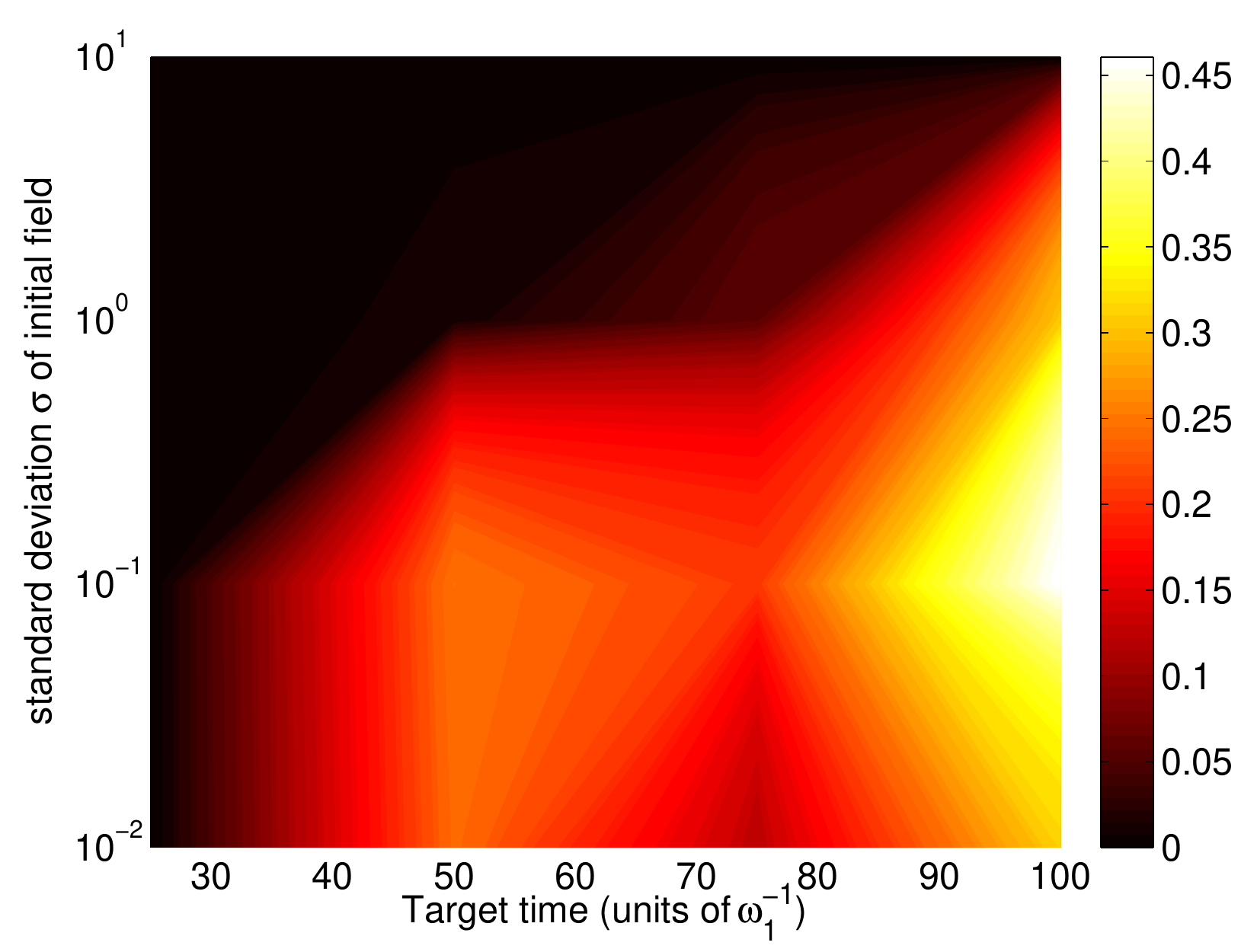}}

\subfloat[\sf Two-qubit CNOT gate (1 noise qubit)]
{\includegraphics[width=0.42\textwidth]{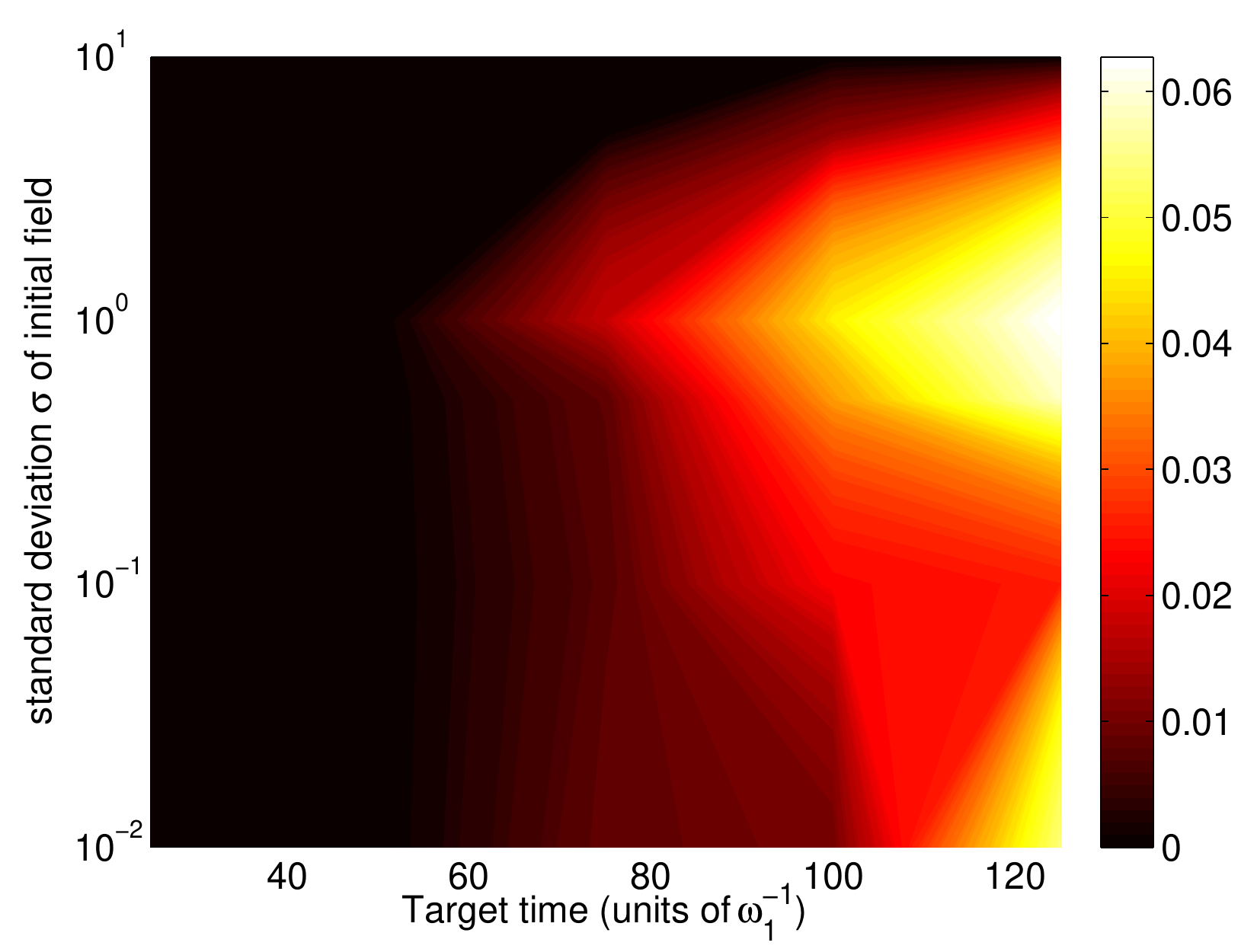}}
\hfill
\subfloat[\sf Two-qubit Id gate (1 noise qubit)]
{\includegraphics[width=0.42\textwidth]{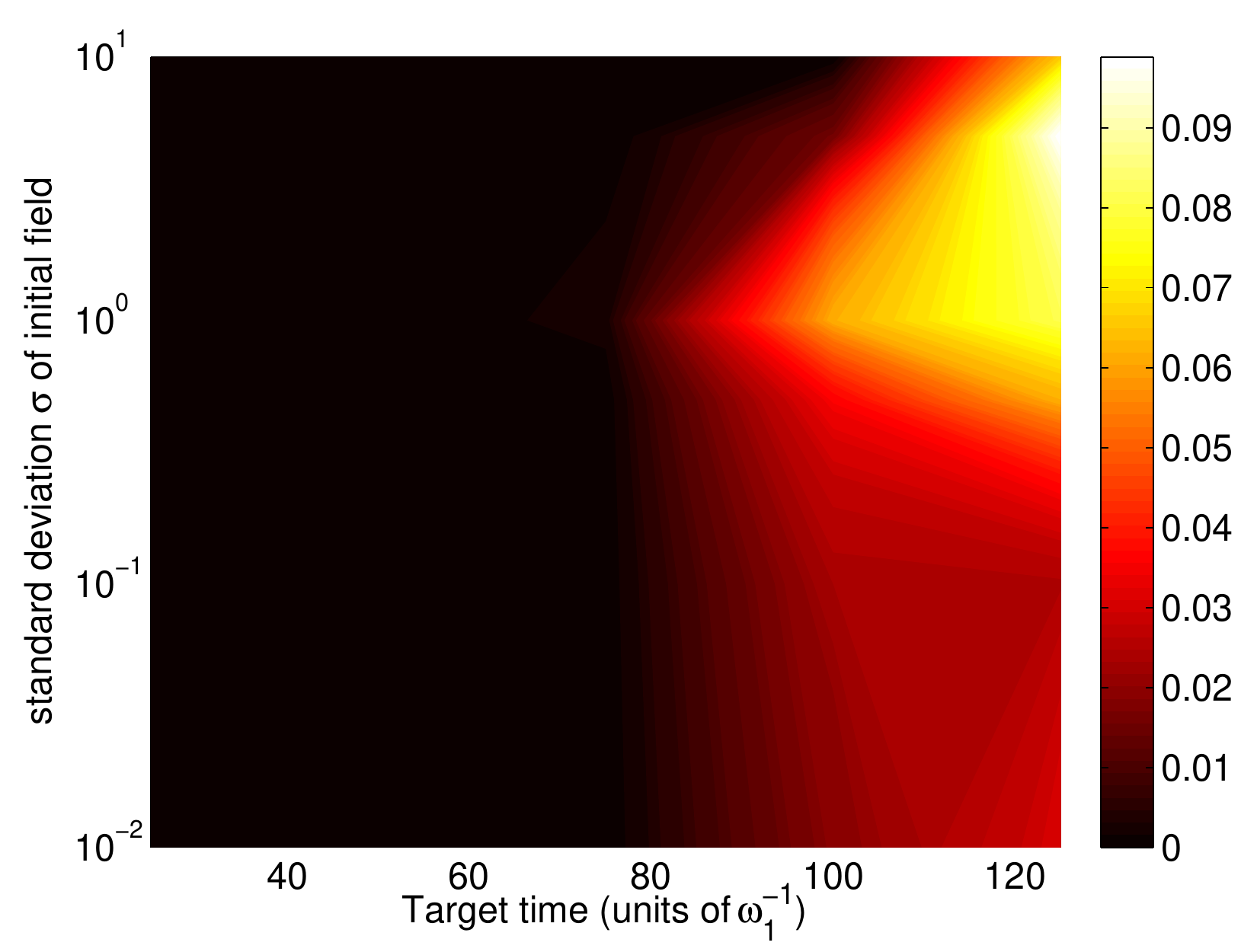}}

\subfloat[\sf Two-qubit CNOT gate (2 noise qubits)]
{\includegraphics[width=0.42\textwidth]{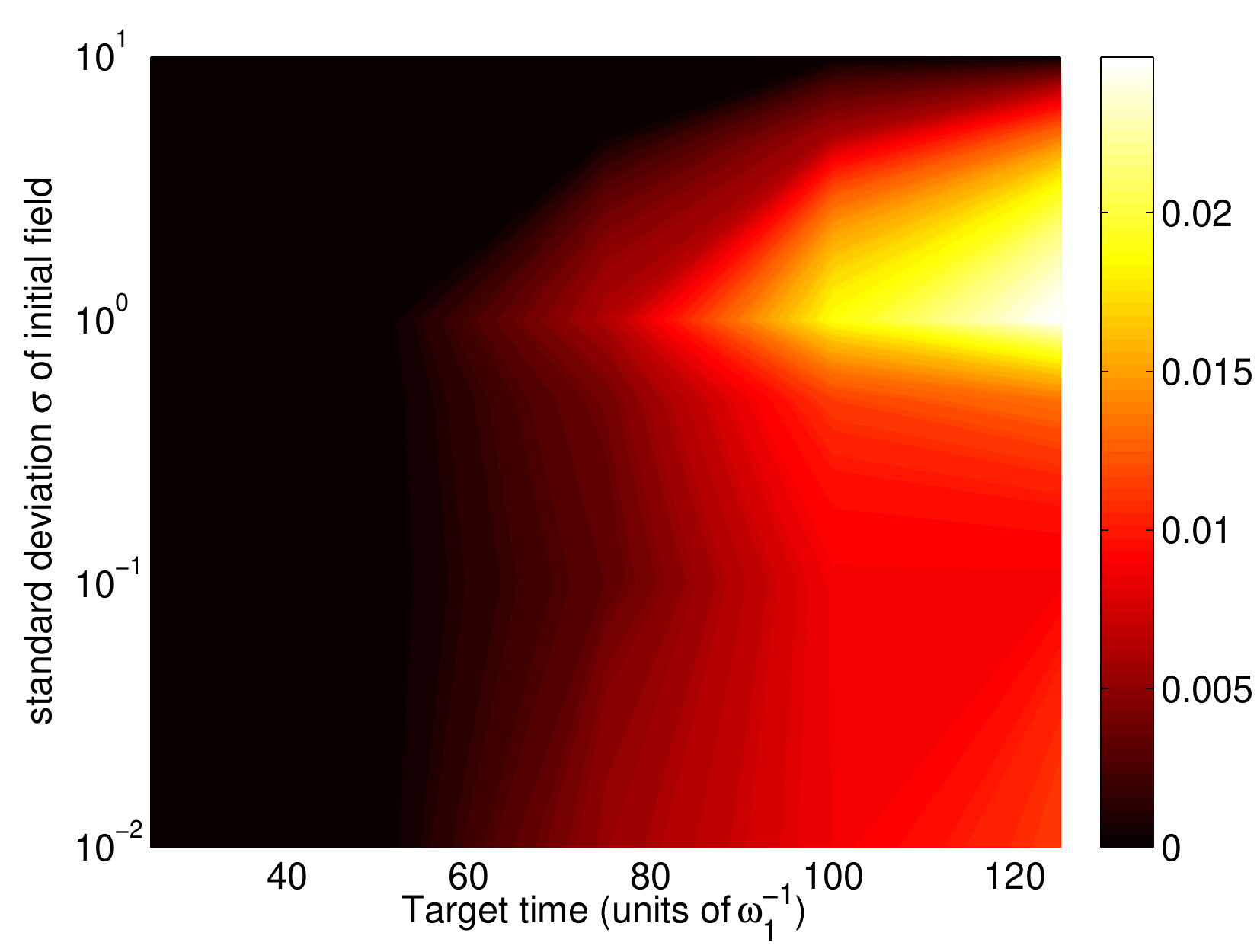}}
\hfill
\subfloat[\sf Two-qubit Id gate (2 noise qubits)]
{\includegraphics[width=0.42\textwidth]{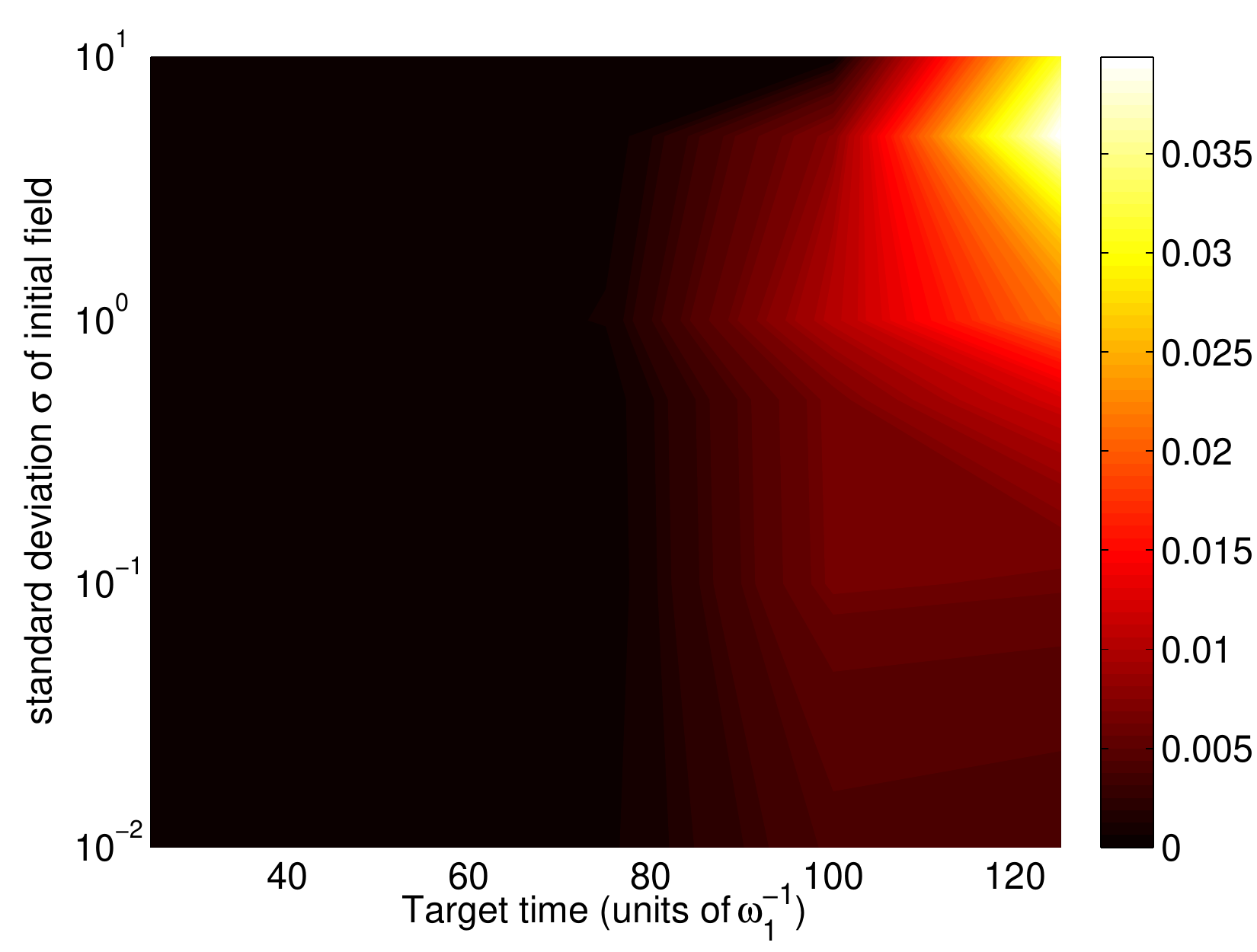}}

\subfloat[\sf Two-qubit CNOT gate (4 noise qubits)]
{\includegraphics[width=0.42\textwidth]{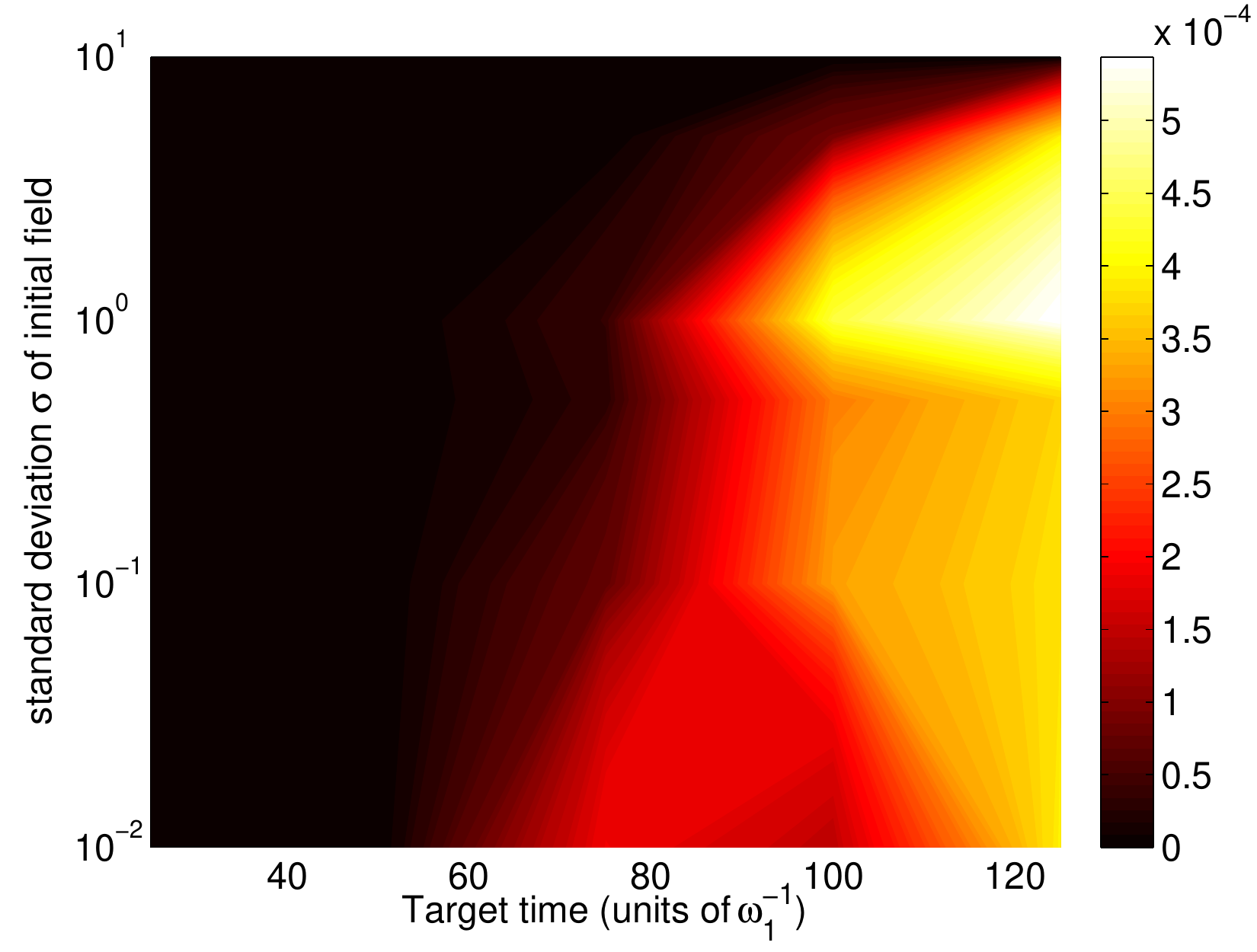}}
\hfill
\subfloat[\sf Two-qubit Id gate (4 noise qubits)]
{\includegraphics[width=0.42\textwidth]{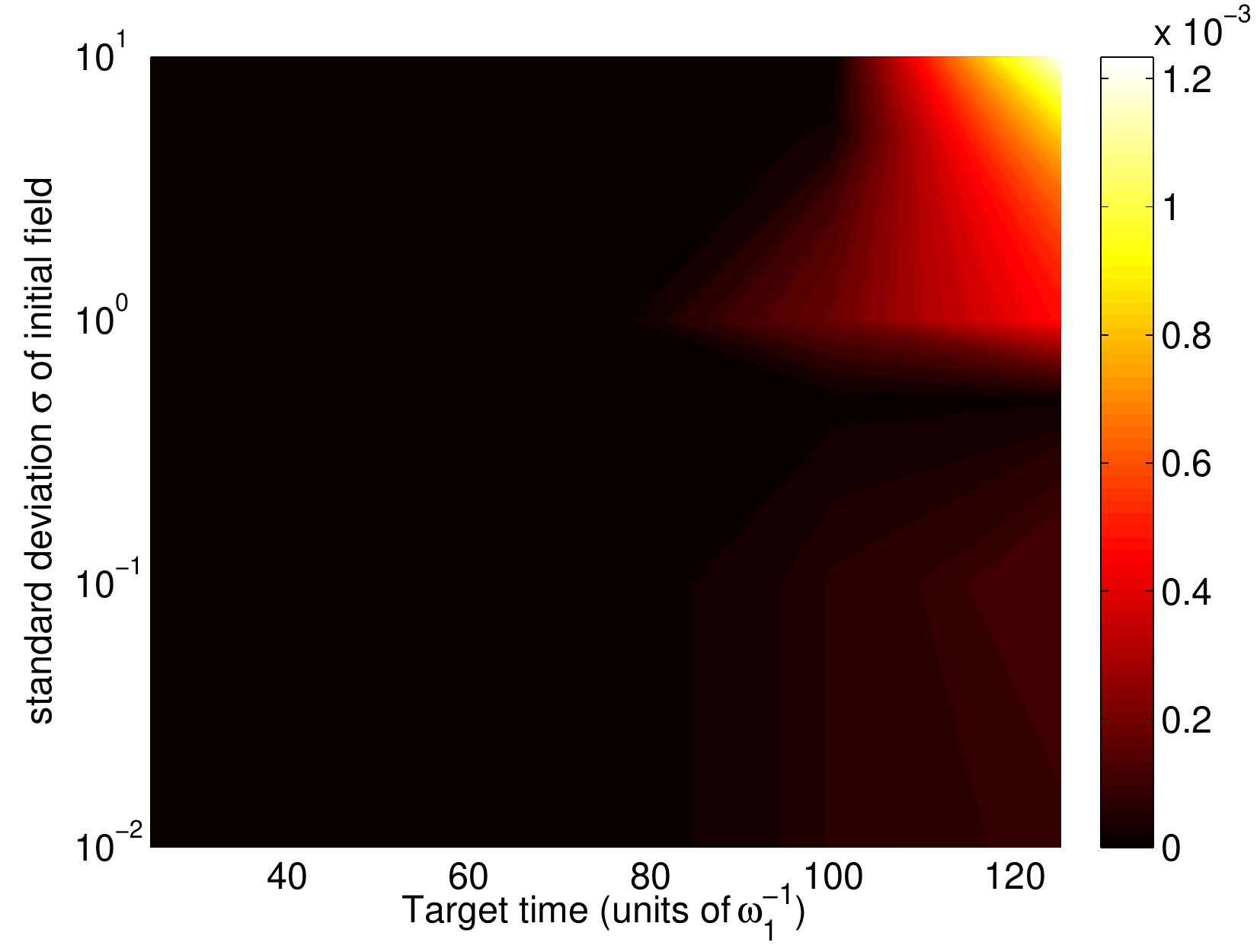}}

\subfloat[\sf CNOT gate, $z$-dephasing]
{\includegraphics[width=0.42\textwidth]{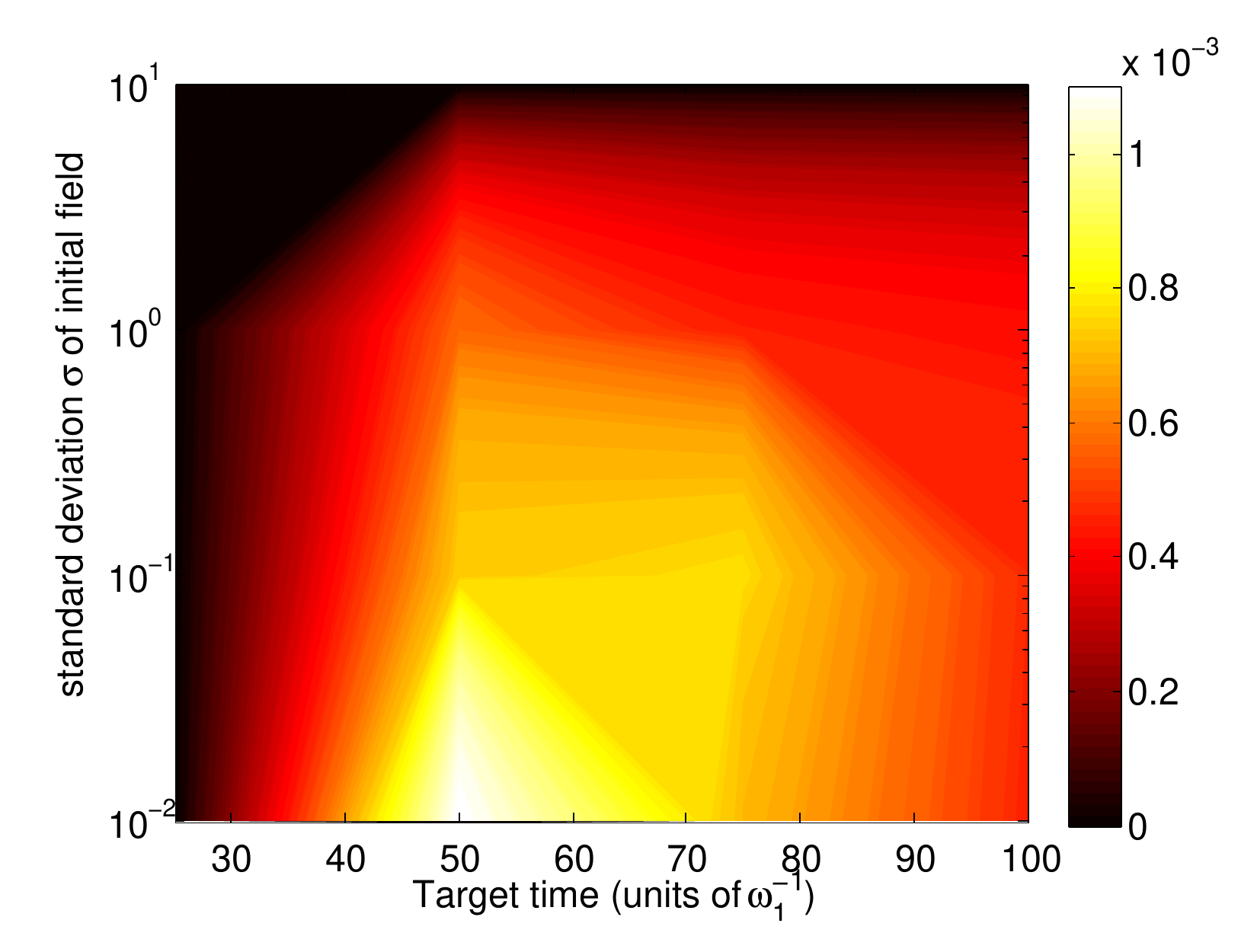}} 
\hfill
\subfloat[\sf Two-qubit Id gate, $z$-dephasing]
{\includegraphics[width=0.42\textwidth]{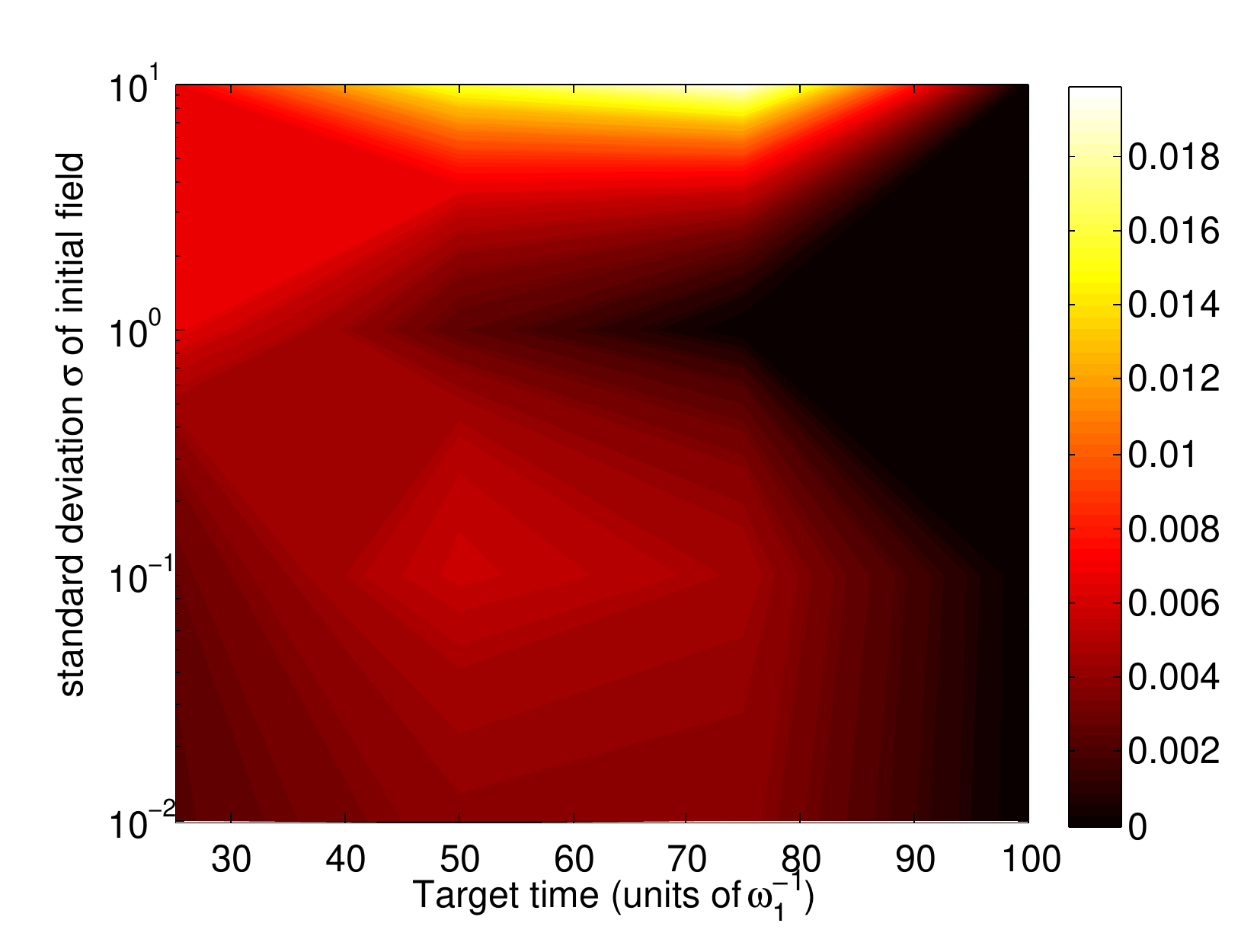}} 
\end{figure*}

For Markovian systems one expects that there exists an optimum time to
implement a desired gate with the highest possible fidelity.  This time
is expected to depend on the details of the system and the Lindblad
operators describing the effects of the bath.  For shorter gate operation
times, the fidelities will be lower due to a lack of time to implement the
required control~\cite{31}; for longer times, decoherence effects
will reduce the fidelity.  For non-Markovian systems, there is generally
no upper limit for the gate operation times and we expect longer times
to improve both success rates and speeds.  In fact, with a sufficiently
long evolution time and no restrictions on the control fields, one can
in theory always achieve perfect fidelities if the composite system
evolves unitarily and is controllable, as is the case for our systems.

The success rate plots (Fig.~\ref{fig:success-rate}) for both a CNOT and
two-qubit identity gate do indeed suggest that there is a lower bound on
the target time but no upper bound.  Moreover, the success rate plots
for one, two, and four noise qubits are very similar and the threshold
value for the gate operation time increases only marginally when more
noise qubits are added, especially for the CNOT gate.  This suggests
that neither the difficulty of finding a control nor the time required
to implement the gate increases substantially when more noise qubits are
added, a very desirable situation in practice.  The picture is slightly
less favourable for the identity gate in that the minimum time necessary
to successfully implement the gate with the desired fidelity increases
more significantly (from 40 in the absence of noise qubits to about 80
for one, two, or four noise qubits).  In the Markovian case,
Fig.~\ref{fig:success-rate}(i) does show a slight decrease in the
success rate for larger gate operation times for the CNOT gate, as
expected based on the observation above. Nevertheless, much more
interesting is the success rate plot for the identity
Fig.~\ref{fig:success-rate}(j), for which there appears to be no minimum
gate operation time but there is a sharp drop in the success rate when
$\tf$ gets too large.

For both models, the success rate plots for the CNOT gate show a sharp
drop in the success probability regardless of the target time when the
amplitudes of the initial fields get too large.  The success speed plots
(Fig.\ref{fig:success-speed}) further suggest the existence of a sweet
spot (white) with regard to the field amplitudes for which we can expect
rapid convergence.  For larger initial field amplitudes, we can expect
poor performance due to trapping in sub-global extrema; for too
small initial field amplitudes, we can expect slow convergence.  It is
therefore desirable to determine the optimum range of initial field
amplitudes at the outset, which could be done by adapting the methodology
which has been successful for closed systems~\cite{32}.  In the
non-Markovian setting the success speed plots for both the CNOT and
identity gate are not too dissimilar; in particular, the success speed
plots have a sweet spot for larger target times around a certain optimum
value of the initial field amplitudes.  In the Markovian setting, on
the other hand, the shortest time to success was achieved for initial
fields with a small magnitude for the CNOT gate but large magnitude for the
identity gate, showing that the optimum initial field amplitudes may
depend on the target gate to be implemented.  In general, the identity
gate was harder to implement with high fidelity than the CNOT.
A given realization of a basic quantum information processor may thus be
more effective in carrying out certain operations than others, and
implementing a trivial gate may, in fact, be harder than implementing a
maximally entangling gate.

\subsection{Pre-optimised fields}

\begin{figure*}[t]
\caption{Performance of pre-optimised fields and improvement when 
environment is taken into account.}  \label{fig:Refine}

\subfloat[\sf 2~qubits, indep.~$z$-dephasing, CNOT, $\tf=75$~$\omega_1^{-1}$]
{\includegraphics[width=0.64\textwidth]{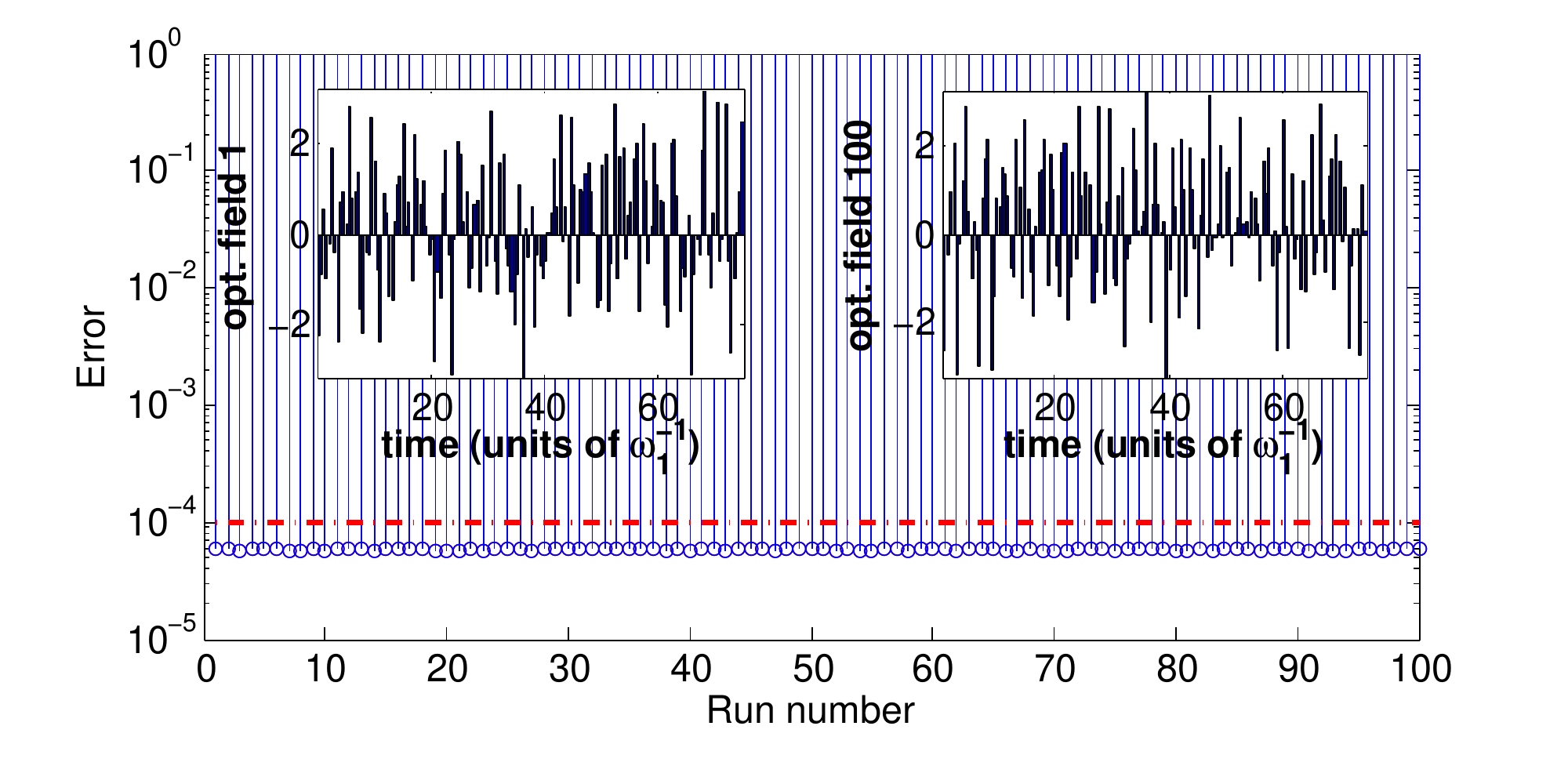}}
\subfloat[\sf 2~qubits, 4 noise qubits, CNOT, $\tf=75$~$\omega_1^{-1}$]
{\includegraphics[width=0.4\textwidth]{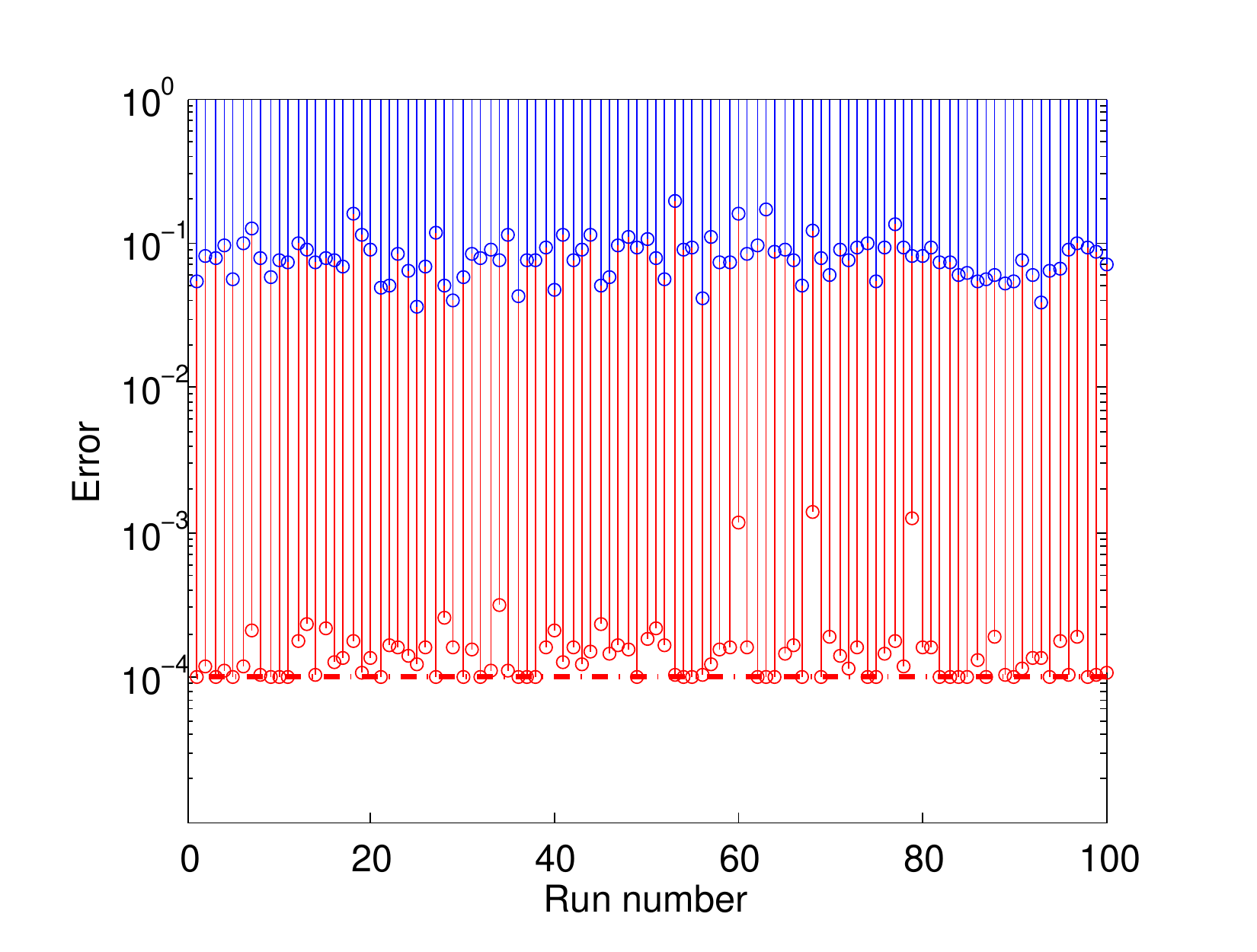}}
\end{figure*}

Another interesting question is how important it is to take the effect
of the environment into account when optimising the gate performance.
In general, one expects a better performance when the environment is taken
into account but this is computationally much more expensive.
It is therefore informative to compare the performance of the fields
optimised with and without taking the environment into account.  To do
this, we first optimise the controls neglecting decoherence and then
start the iteration in the dissipative case with an optimal control
obtained for the Hamiltonian case.  Here, we observe major differences
depending on the type of environment.  

In the Markovian case the optimal fields obtained without the
environment performed quite well --- taking the decoherence mechanisms
into account resulted in little, if any, improvement, as shown in
Fig.~\ref{fig:Refine}(a).  The blue stems in the figure indicate the
error achieved when the pre-optimised fields are directly applied to the
dissipative system, while the red stems indicate the final errors
achieved when these fields are used as inputs for optimisation with
decoherence.  The red and blue stems are virtually indistinguishable
which suggests that the performance of the optimal control fields
obtained without considering decoherence is comparable to that obtained
taking the Markovian environment into account.  This was the case across
the board for thousands of simulations for all of the systems, gates,
and Markovian environments considered here.  This disappointing
performance of optimal control for Markovian systems may seem surprising
in view of some earlier work but it is actually in line with results in
\cite{42} as substantial improvements in the fidelities achieved with
optimal control in this work were observed only for a qubit encoded in a
weakly relaxing subspace, a distinctly different situation from the one
considered here.  In general, the simulation results are consistent with
the observations that open-loop coherent control cannot undo the effects
of Markovian decoherence~\cite{allan}.  This should not be interpreted
to imply that optimal control design is not useful for systems subject
to Markovian decoherence.  Optimal control generally still leads to more
efficient quantum gates; however, there are fundamental limitations and
it may be necessary to combine optimal control design with other
strategies such as robust encoding of quantum information or feedback.

\begin{figure*}
\caption{Bloch sphere trajectory plots for 1 system qubit and 4 noise qubits.}
\label{fig:Bloch}
\vspace{-2ex}
\subfloat[\sf Field optimised not taking noise qubits into account leads
 to unwanted excitation of noise qubits reducing the
 fidelity.]{\includegraphics[width=\textwidth]{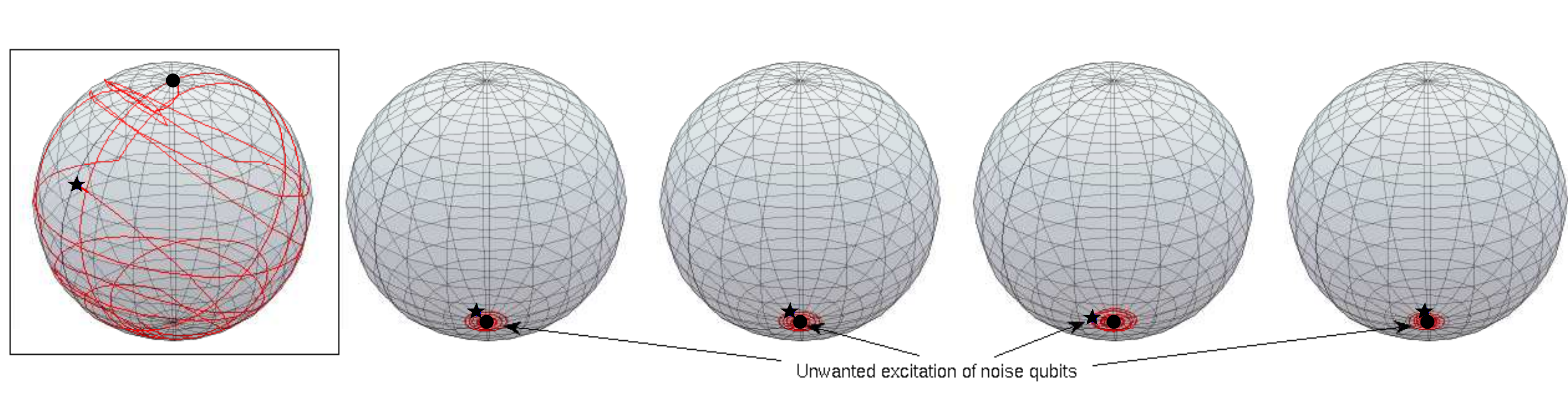}}
 \\[2ex] \subfloat[\sf Field optimized taking noise qubits into account
 suppresses excitation of noise qubits restoring
 fidelity.]{\includegraphics[width=\textwidth]{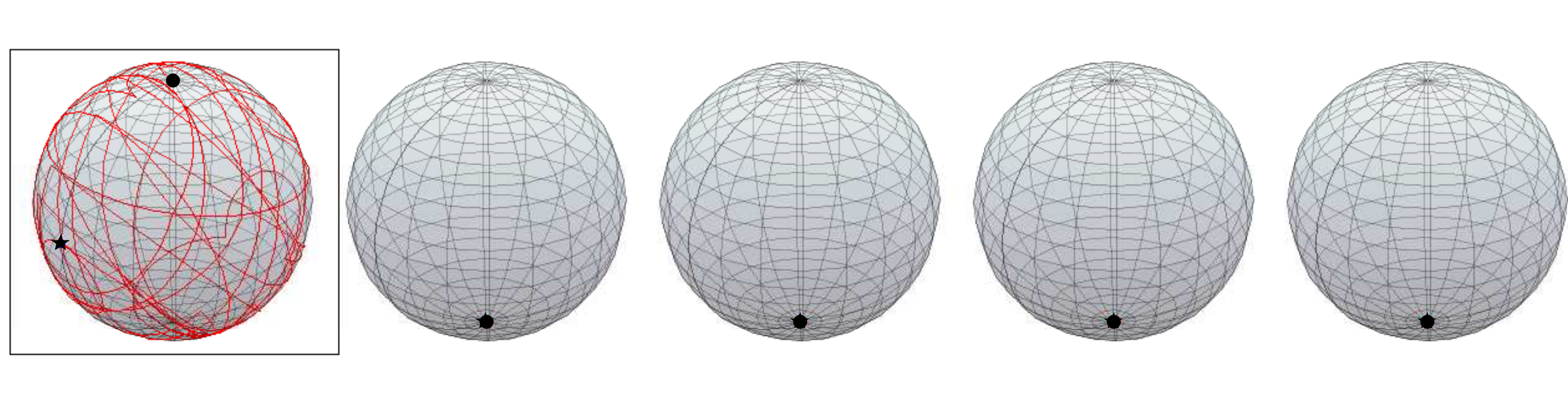}}
 \\ \subfloat[\sf Field optimised taking noise qubits into account fails
 to produce desired dynamics when noise qubits not shielded from
 field.]{\includegraphics[width=\textwidth]{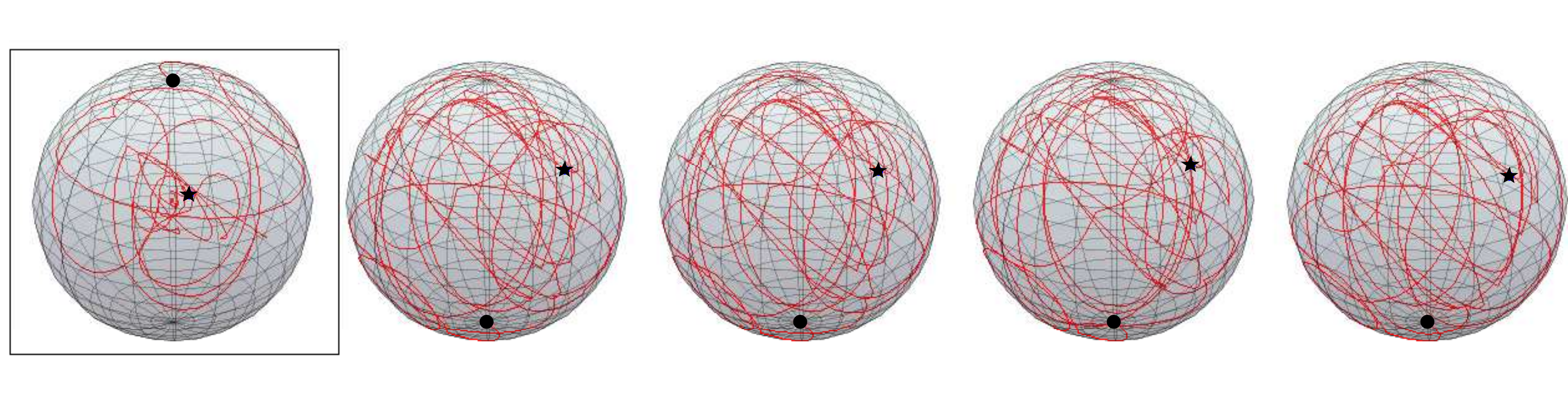}}
 \\ \subfloat[\sf Field optimised taking excitation of noise qubits into
 account can restore high fidelities by exploiting coherence
 revivals.]{\includegraphics[width=\textwidth]{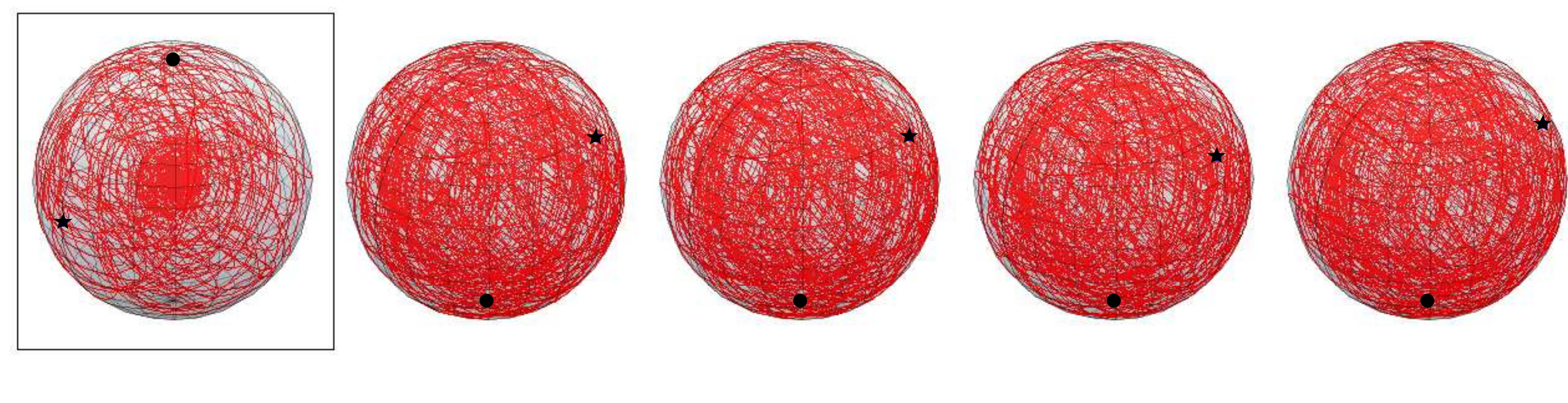}}
\\
Initial and final states indicated by $\bullet$ and $\star$,
respectively.  First qubit (framed) is the system qubit.  The gate
operation time is $\tf=25$~$\omega_1^{-1}$ in the first three cases and
$\tf=314$~$\omega_1^{-1}$ in the fourth case.
\end{figure*}

In the non-Markovian case, pre-optimised fields almost always performed
poorly and the fidelities could be considerably improved by taking the
noise qubits into account in the optimisation, as illustrated in
Fig.~\ref{fig:Refine}(b).  Note that we observe a reduction in the error
by about three orders of magnitude.  Starting with pre-optimised fields
also did not improve the convergence speed or increase the attainable
fidelities, compared to starting with random initial fields, suggesting
that pre-optimisation is not beneficial in this case.  The reason why
the fields obtained without taking the noise qubits into account perform
poorly can be seen if we plot the evolution of the Bloch vectors of the
system and noise qubits.  Fig.~\ref{fig:Bloch}(a) shows the evolution of
a single system qubit and four noise qubits under a field optimised to
implement a HAD gate on the system qubit without taking the noise qubits
into account.  We see that although the field does not excite the noise
qubits directly, they are indirectly excited due to the coupling to the
system qubit. In addition, although the magnitude of the excitation is
much less than that of the system qubit, it is sufficient to create
entanglement between the system and noise qubits, reducing the length of
the Bloch vector of the system qubit and the gate fidelity.  When the
noise qubits are taken into account in the optimisation, the algorithm
finds a field that avoids indirect excitation of the noise qubits almost
entirely, as shown in Fig.~\ref{fig:Bloch}(b).  Entanglement between the
system and noise qubits in this case is negligible and the gate fidelity
is restored.

\subsection{Control Mechanisms and Effects of Field Leakage}

\begin{figure*}
\caption{Effect of Field Leakage}
\label{fig:Leakage}

\subfloat[\sf HAD Gate error vs gate operation time $\tf$]{
\includegraphics[width=0.49\textwidth]{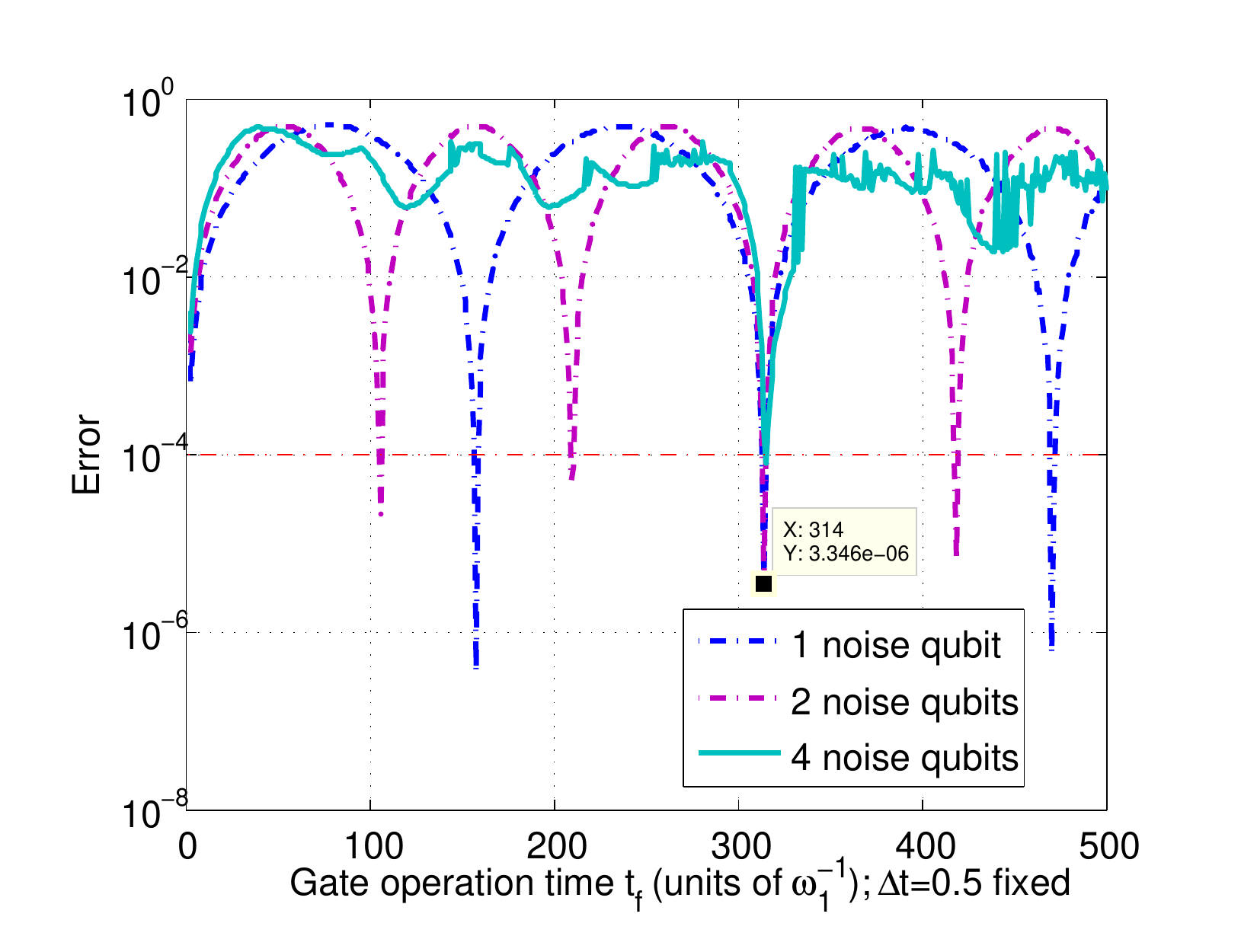}}
\subfloat[\sf Convergence for HAD gate with two noise qubits]{
\includegraphics[width=0.49\textwidth]{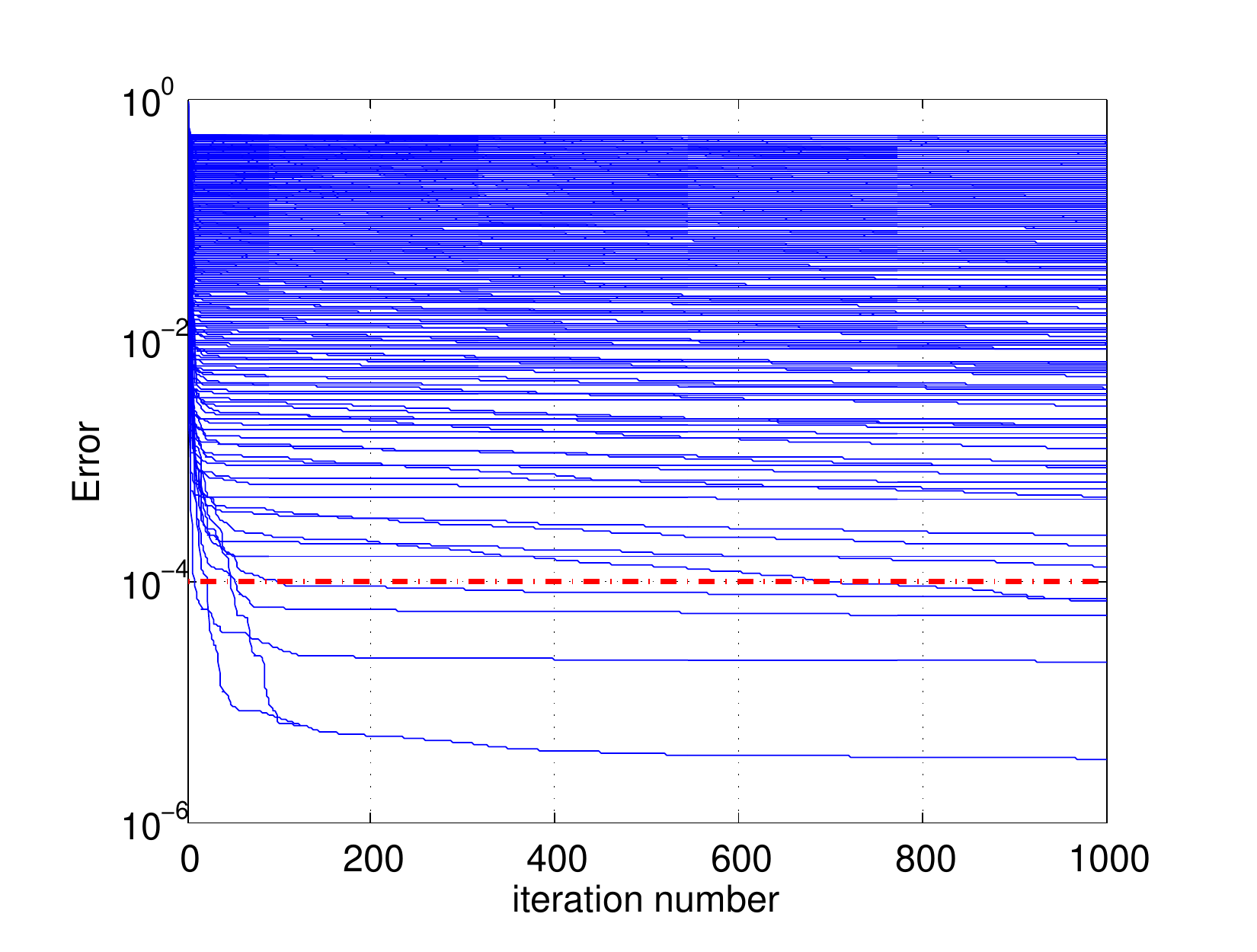}}

\subfloat[\sf Gate error for 2-qubit system with standard noise
qubit]{
\includegraphics[width=0.49\textwidth]{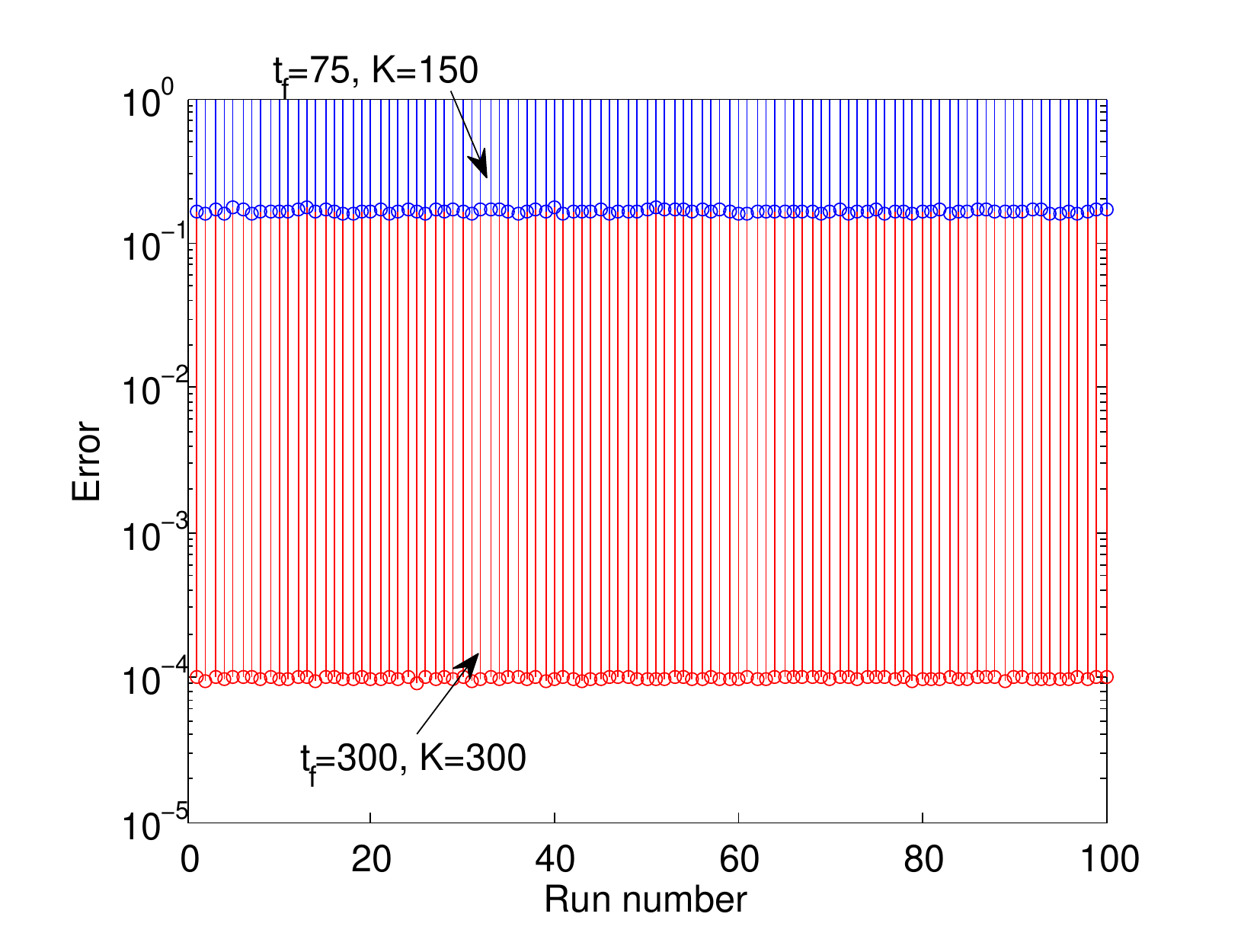}}
\subfloat[\sf Gate error for 2-qubit system with far-detuned
noise qubit]{\includegraphics[width=0.49\textwidth]{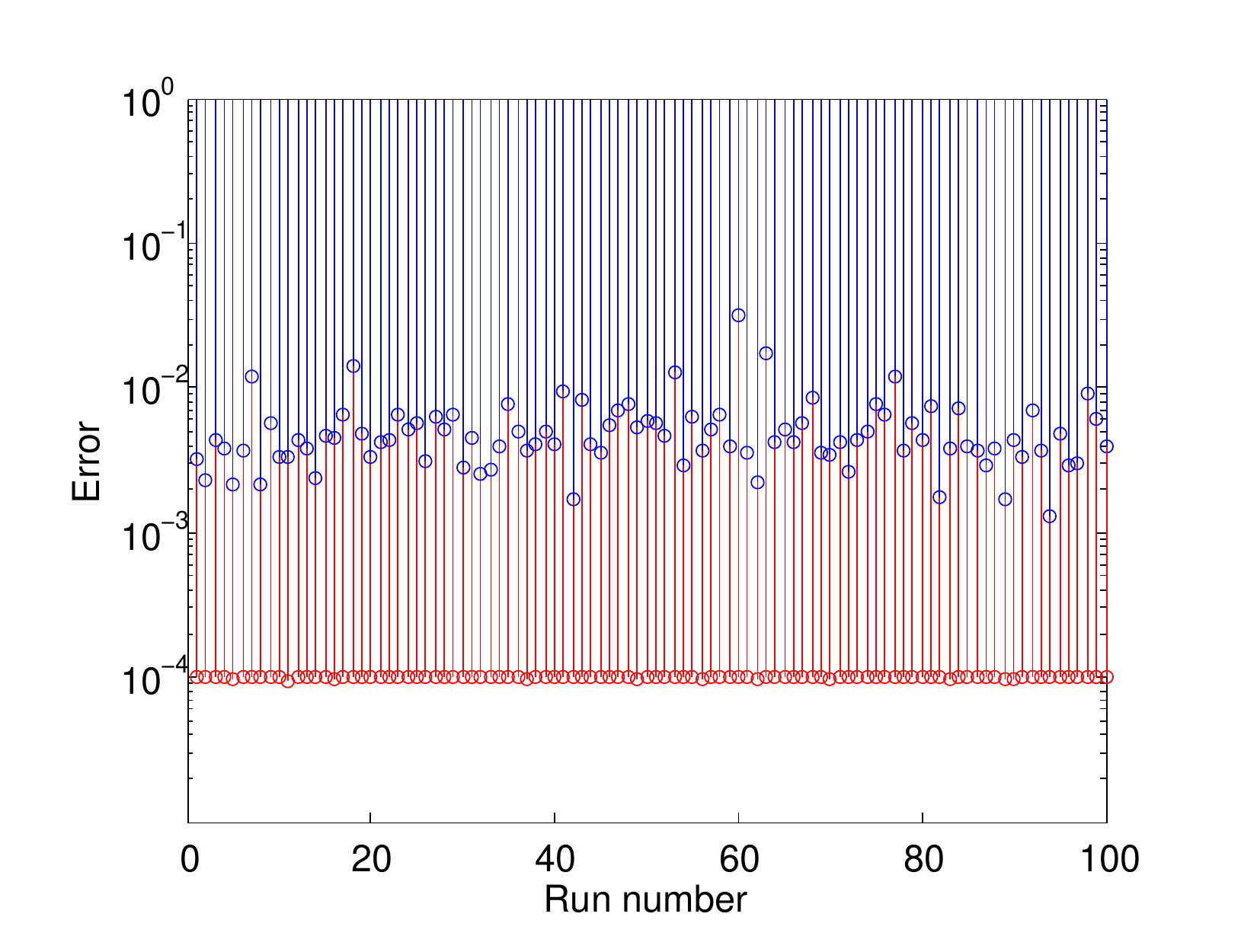}}
\end{figure*}

In the previous calculations it was assumed that the noise qubits are
shielded from the control field and that the field does not affect the
environment in any way, an assumption that is not necessarily realistic.
To assess the effect of field leakage we considered what happened if the
noise qubits were affected by the fields as well --- for example, due to field
leakage.  In the worst case scenario, the field seen by nearby noise
qubits will be similar to that seen by the system qubits and they may
interact with the field in a similar way to the system qubits.  In this
case, the final fidelities significantly decreased.  In fact, the optimal
fields obtained without considering the effect of the field on the noise
qubits proved useless, as shown in the Bloch trajectory plot in figure
\ref{fig:Bloch}(c).  The noise qubits are now excited not only weakly by
the coupling to the system qubit but also directly by the control field,
resulting in significant excitation and entanglement between the system and
noise qubits, evident by shrinkage of the length of the Bloch vector 
(especially for the system qubit).  Nevertheless, if leakage is taken
into account in the optimisation, the results can be improved, as shown in
Fig.~\ref{fig:Bloch}(d).  We see that unlike in the no-leakage case (b),
both the system and noise qubits are excited and now undergo complex
evolutions.  There is intermittent entanglement and the noise qubits are
left in excited states at the final time; still, the field is such that
the system qubit is disentangled from the noise qubits at the final time
and the desired Hadamard gate is implemented.

However, to achieve such a performance, the gate operation time had to
be substantially increased from $\tf=25$ to $\tf=314$ and chosen very
carefully.  When the target times and time resolution of the field were
kept the same as before, the results were poor.
Fig.~\ref{fig:Leakage}(a) illustrates the complicated dependence of the
attainable gate fidelities on the gate operation time for a single
system qubit surrounded by multiple noise qubits with frequencies very
close to the system qubit's frequency.  Only a few runs in
Fig.~\ref{fig:Leakage}(b), corresponding to the dips seen in
Fig.~\ref{fig:Leakage}(a), reach the error threshold of $10^{-4}$ but
these successful runs usually reach it in relatively few iterations.
Due to the rapid flattening out of the trajectories, most runs could
have been terminated after fewer iterations as it is evident that they
will not attain high fidelities.  The significant increase in the amount
of time needed to implement a simple single qubit gate can be explained
in terms of global control.  When the field leakage is taken into
account, the control problem becomes a global control problem for a five
qubit system and the only way we can achieve selective excitation of a
single qubit is by exploiting the differences in the resonance
frequencies of the qubits.  Since frequency differences between the
system and noise qubits are small in our case, more time is required to
achieve selectivity.  This suggests that the effect of leakage should be
significantly reduced if the resonance frequencies of the noise qubits
are significantly different from those of the system qubits.

To test this latter idea, we performed additional runs for a two-qubit
system with one standard noise qubit, and the same system with a 
far-detuned noise qubit with $\Delta \omega=5$ instead.  For our
standard noise qubit, Fig.~\ref{fig:Leakage}(c) shows that the gate
errors for the two-qubit CNOT gate for $\tf=75$ and $K=150$ are large
(blue stems).  The gate operation times and time resolution of the field
need to be increased substantially to achieve gate errors below
$10^{-4}$ (red stems).  However, Fig.~\ref{fig:Leakage}(d) demonstrates
that for a far-detuned noise qubit all 100 runs for $\tf=75$ and $K=150$
(and standard deviation of the initial fields equal to 1) reached the
error threshold of $10^{-4}$ even when the noise qubit was not shielded,
provided the effects of the fields on the noise qubit were taken into
account in the optimisation.  Accordingly, shielding the noise qubits
from the control field may not be necessary, provided they are
sufficiently detuned from the system qubit frequencies.  This suggests
that increasing the detuning of the noise qubits from the system qubits
could be one way to mitigate the effect of the control fields on the
noise qubits if leakage is unavoidable.  For instance, this could be
achieved via control electrodes~\cite{28} acting on the noise qubits
(although in practice such direct control of the noise qubits could be
technologically challenging).  In this case, if the noise qubit
frequencies are too similar to the system qubit frequencies and they
cannot be shielded, high fidelities and good gate operation times may be
out of reach even with the best optimal control.

\subsection{Characteristics of the Optimal Fields}

While finding solutions that achieve high fidelities is crucial and
optimising the algorithms and parameters to achieve rapid convergence is
highly desirable, not all optimal control solutions are created equal.
In practice, the feasibility of the implementation of a control depends
on its basic characteristics such as its pulse energy, amplitude, and
bandwidth.  Generally, fields with lower amplitudes and pulse energies
are desirable as strong fields can be more difficult to implement and
cause unwanted excitations or may even break the model the optimisation
is based on.

\begin{figure*}
\caption{Distribution of maximum amplitudes of optimal fields for
 different initial field amplitudes: $\delta=0.1$ (blue), $1$ (red) and
 $10$ (green) {\sf ($T$ in units of $\omega_1^{-1}$)}}  
\label{fig:Hist} 

\subfloat[\sf Had gate, spontaneous emission, $\tf=25$]
{\includegraphics[width=0.49\textwidth]{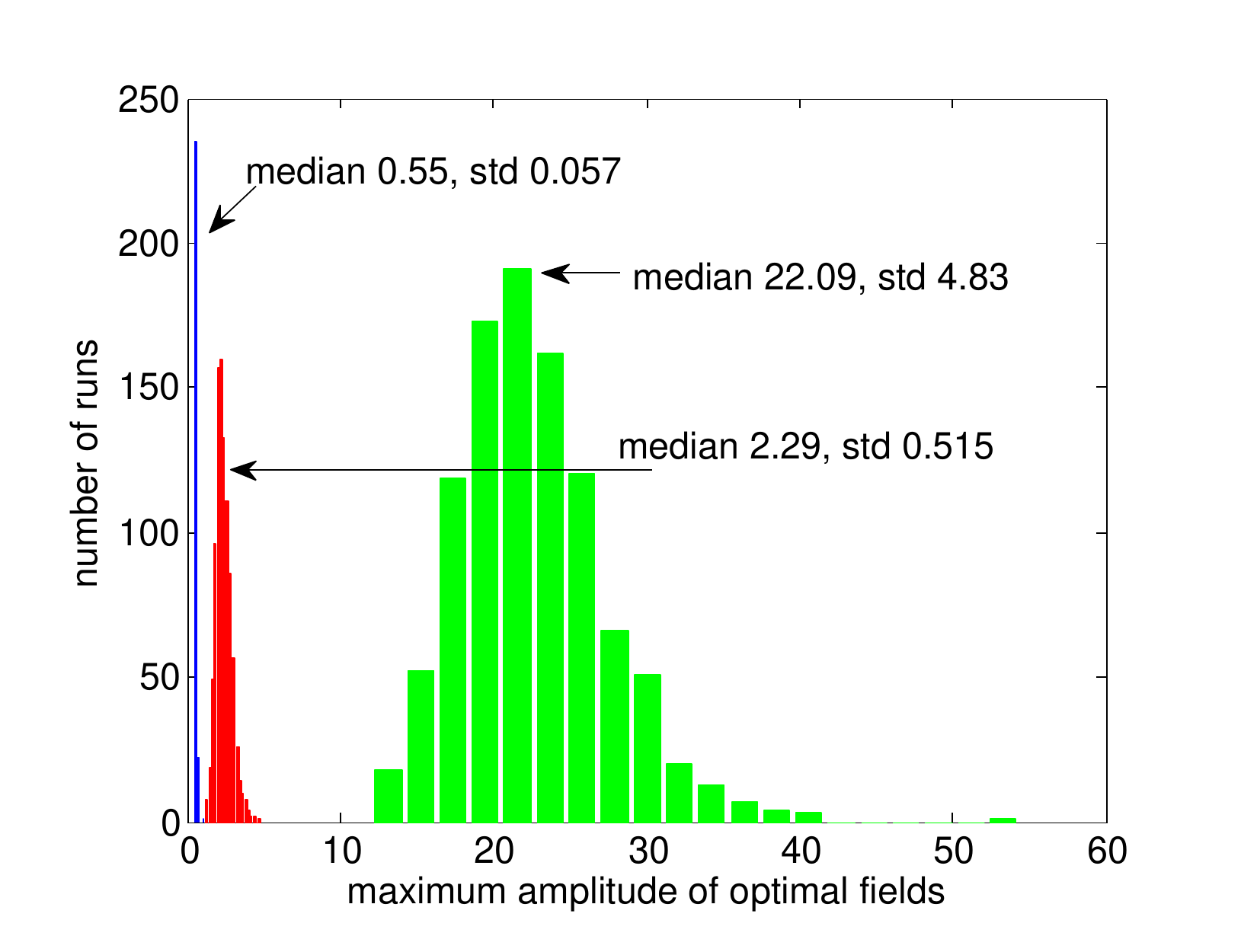}}
\subfloat[\sf Id gate, spontaneous emission, $\tf=25$]
{\includegraphics[width=0.49\textwidth]{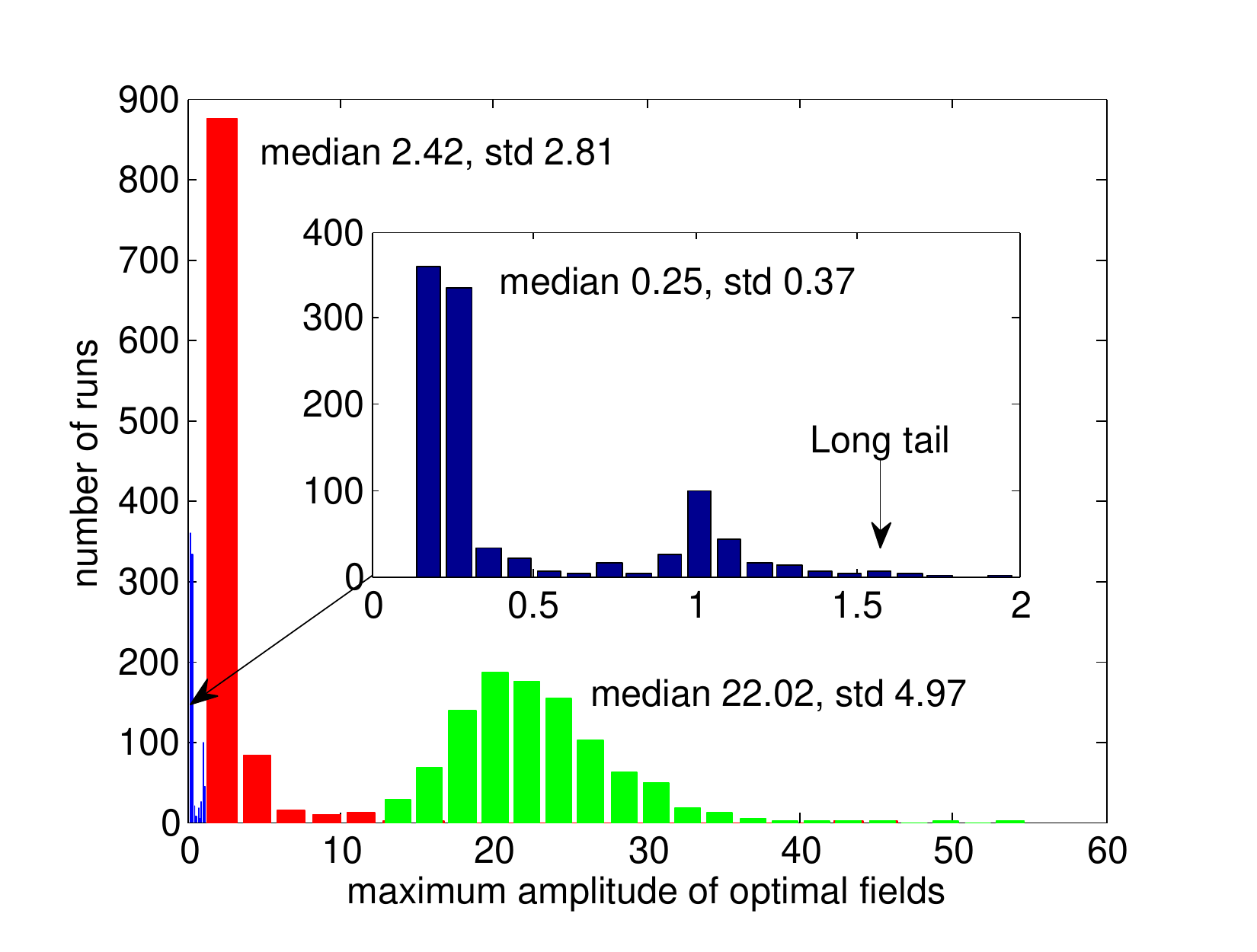}}
\\[2ex]
\subfloat[\sf Had gate, 1 qubit, 2 noise qubits, $\tf=25$]
{\includegraphics[width=0.49\textwidth]{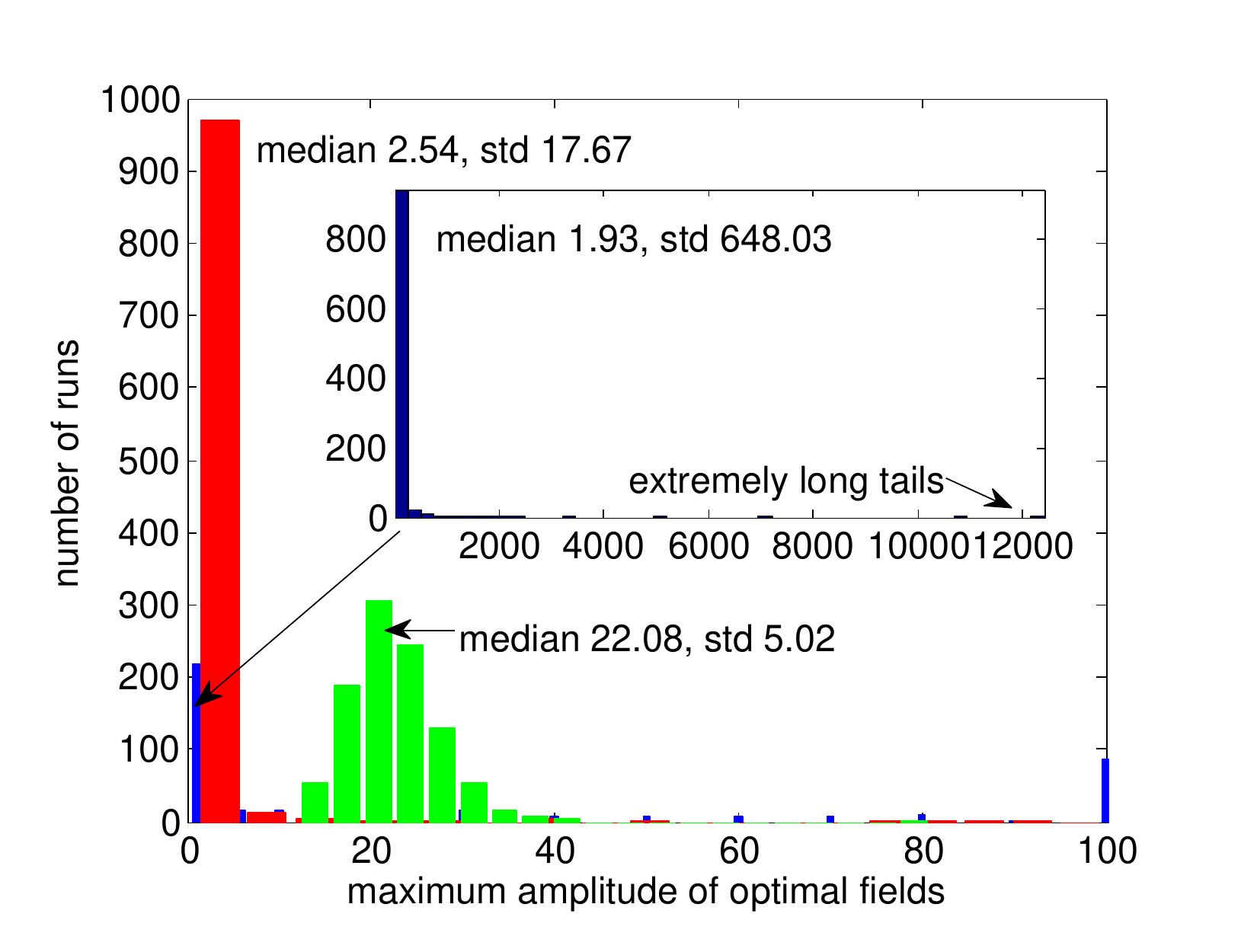}} 
\subfloat[\sf Had gate, 1 qubit, 2 noise qubits, $\tf=3$]
{\includegraphics[width=0.49\textwidth]{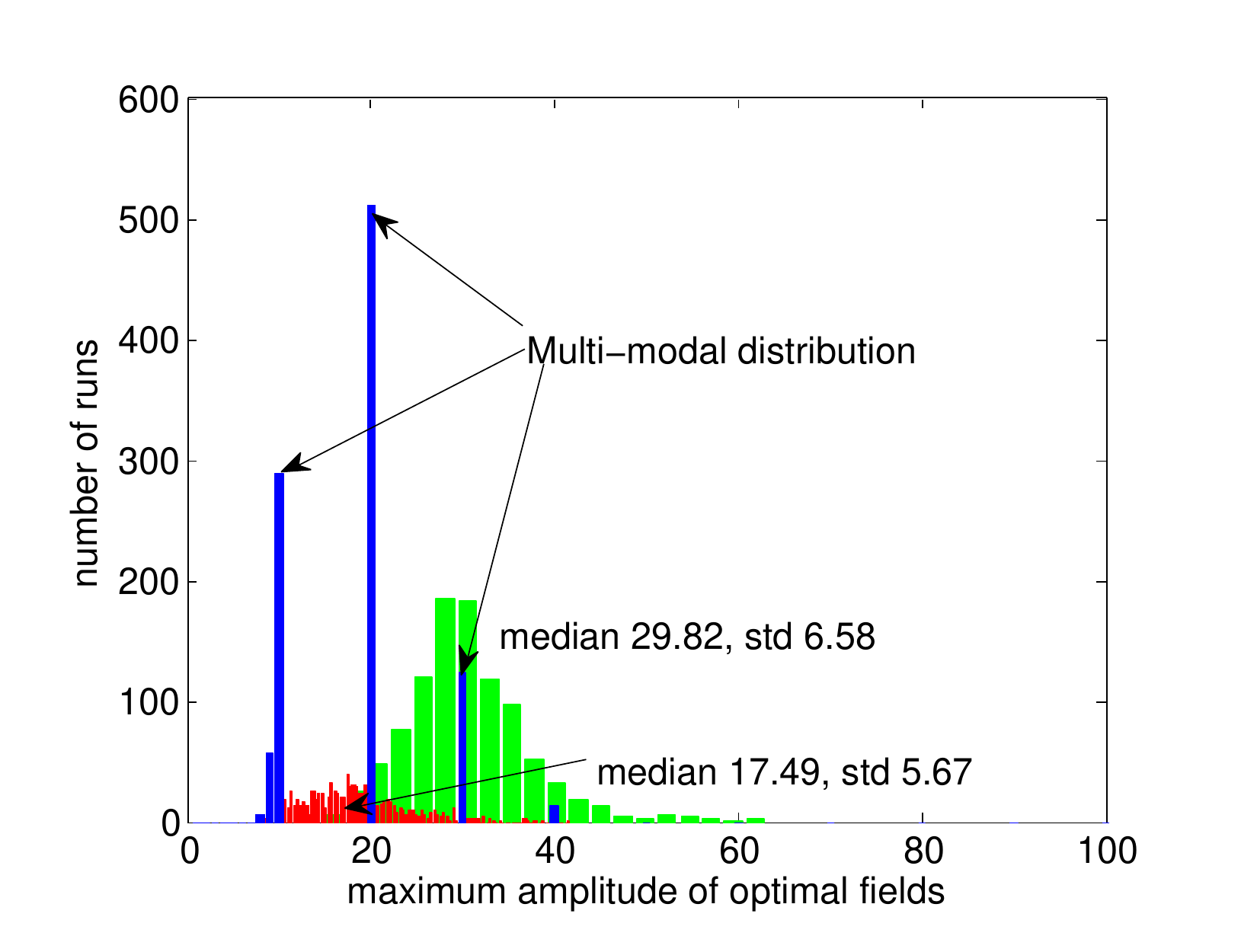}} 
\end{figure*}

In the Markovian case, analysis of the maximum field amplitudes of a
large number of successful runs, that is, runs that achieved gate errors
$<10^{-4}$, revealed a strong correlation with the magnitude of the initial
fields, as determined by the standard deviation $\delta$ of the Gaussian
distribution we sample from. This can be seen in the histogram plots of the
maximum field amplitudes for 100 successful runs, for different values of
the standard deviation of the initial fields (Fig.~\ref{fig:Hist}).  In
particular, low-amplitude initial fields ($\delta$ small) produce a
narrow distribution with a small median value, meaning we are likely to
converge to a low-amplitude field if we start with an initial field with a
small amplitude.  Accordingly, there is no need for the addition of
penalty terms to the objective functional to constrain the field
amplitudes.  Penalty terms are generally undesirable as they prevent the
algorithm from converging to a global optimum of the objective function
and lead to slower convergence.  In some cases --- for example,
Fig.~\ref{fig:Hist}(b) --- the distribution has a long tail, that is,
there are some runs starting with small fields that nonetheless converge
to high-amplitude solutions.  However, the probability of this happening
is generally small, and if we do converge to a field with unexpectedly
large amplitudes, we can most likely find a better solution by repeating
the unconstrained optimisation with another suitably low-amplitude
initial field.

In the non-Markovian case, we observe similar behaviour but the maximum
field amplitude histograms tend to be less sharply peaked than in the
Markovian case and can exhibit multi-modal features as in
Fig.~\ref{fig:Hist}(d).  In some cases, we also observe scattered
outliers, which are fields with maximum amplitudes orders of
magnitude greater than the median of the distribution, as shown in
Fig.~\ref{fig:Hist}(c).  In this setting, there can be substantial
differences in the amplitudes of the optimal fields for initial
fields with the same standard deviation; furthermore, the maximum amplitudes
and energies of the optimal fields can be substantially reduced by
running the algorithm several times to find a solution belonging to the
lowest amplitude peak.

\begin{figure*}
\caption{Scatter plot of the fluences of the converged fields versus the
 fluences of the initial fields for a two-qubit CNOT gate with 4 noise
 qubits.} \label{fig:scatter}
 \center\includegraphics[width=0.49\textwidth]{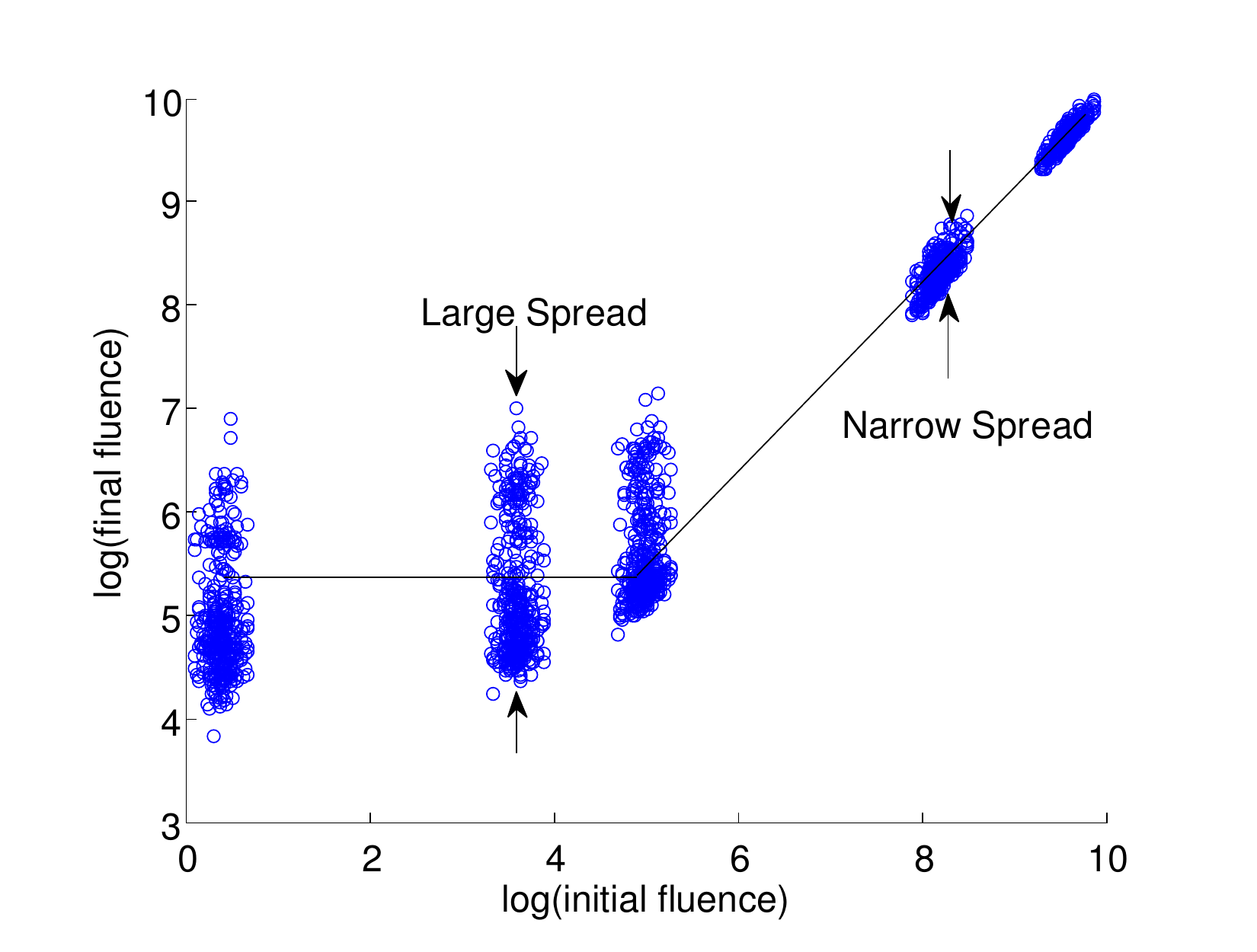}
\end{figure*}

We also studied the correlations between the initial and final field
fluences.  As shown in Fig.~\ref{fig:scatter}, for large initial
fluences, the final fluences cluster around large values that appear to be
proportional to the initial fluence.  For initial fluences below a
certain value, the distribution of final fluences stabilises, so that we
cannot decrease the fluence of the optimal fields below a certain
threshold regardless of how small we choose the initial fluences.  One
way to explain this behaviour is that there is a certain minimum energy
required to achieve the control objective. If we start with initial
fields with energies (fluences) well below this value, the algorithm has
to increase the field energy at least until we reach the minimum energy
required to achieve the control objective. If we start with fields well
below this minimum energy threshold, it may take many iterations to
reach a regime where optimal controls exist, thus increasing the time
required to find a solution.  On the other hand, if we start with
fluences well above the minimum required to achieve the control
objective, the algorithm has no pressure to decrease the field energy --
as we impose no penalty on the field fluences or amplitudes -- and thus
is likely to converge to an undesirable high-energy pulse. There is thus
a best value for the variance of the initial fields to encourage rapid
convergence to a low-energy optimal field.

\section{Summary and Conclusions}

We have presented a general framework for optimal control of quantum
processes for systems in Markovian and non-Markovian environments.  By
suitable discretisation of the fields, we can reduce the complexity of
the simulations and obtain analytic gradient formulae that enable us to
apply efficient gradient-based optimisation algorithms such as second
order quasi-Newton methods.  To better understand the performance of the
algorithms and the influence of parameters such as gate operation times
and initial fields, we introduced new notions of success rates and
speeds and used density plots as a new investigative tool.  Using these
tools combined with convergence and statistical analyses of basic
characteristics of the optimal fields, we have investigated the
reliability and efficiency with which optimally controlled quantum gates
can be implemented on one, two, and three system qubits in Markovian and
non-Markovian models.  Although both models are based on rather
different assumptions, the results for the control optimisation share
some similarities.

First, the success rate and speed plots show that both the target gate
operation times and magnitudes of the initial fields have a significant
impact on the success rates and the expected times to succeed.  For
non-Markovian systems, however, the need to exploit coherence revivals
can lead to a complex dependence of the attainable fidelities on the
target times especially when the noise qubits are not shielded from the
control and have similar frequencies as the system qubits.  The success
rate and speed plots also suggest that starting with too large field
amplitudes increases the likelihood of trapping in many cases, and an
analysis of the amplitude distributions for the optimal fields obtained
shows that it can also lead to convergence to generally undesirable
high-energy optimal fields. Too small amplitudes, on the other hand,
tend to result in a slow convergence.  It is therefore desirable to
determine the optimum range of the initial field amplitudes, based on
knowledge of the actual physical system under consideration~\cite{32}.

Performing several optimisation runs for different initial fields is
advantageous in all regimes considered. In the non-Markovian case, due
to the spread of asymptotic fidelity values, more than one attempt may
be needed to obtain a suitably low error solution. With no environment,
a run which suffers from severe slowdown, even if only intermittently,
can be abandoned in favour of starting a fresh run having a lower
expected time to completion.  In the Markovian case, both of these
reasons apply and it is even more obvious than for non-Markovian systems
that termination conditions are needed to avoid wasting time on the long
tails observed in the convergence plots.  Another reason to perform
several runs is the considerable variability in the average and maximum
amplitudes of the optimal fields across runs. This issue can also be
addressed by imposing constraints during the optimisation procedure, in
contrast to adding penalty terms to the objectives (which should be
avoided since they limit the reachable fidelities~\cite{25}).

However, there are also significant differences in the results for the
two models.  Although optimal control is beneficial to find efficient
solutions in both cases, in the Markovian setting the potential for
optimal control is significantly reduced as the control cannot undo the
effects of the environment, at least not if we are limited to open-loop
coherent control.  In that case, the benefits of optimal control are
mainly the ability to implement operations faster or to reduce the
detrimental effects of the environment by taking advantage of
low-decoherence subspaces (if these exist).  Results showing longer
target times to be detrimental in the Markovian setting, at least in the
absence of low-decoherence subspaces, support this conclusion.  Perhaps
the most surprising result of our simulations, however, is that the
fields optimised without taking the environment into account still
seemed to be optimal in the Markovian setting, at least if the
decoherence is relatively uniform and sufficiently weak to be considered
a perturbation of the Hamiltonian evolution.  This was the case for all
of the thousands of runs performed for single, two-, and three-qubit
gates and various initial conditions.  Furthermore, the simulations
strongly suggest that the precise type of Markovian environment is not
particularly important in that the fields that were optimal for
independent dephasing in the $x$-basis achieved virtually the same
performance for dephasing in the $z$-basis.  Interestingly, independent
relaxation (spontaneous emission) of the same strength was slightly less
detrimental to the gate fidelities while correlated dephasing proved
slightly more detrimental for the multi-qubit gates.  These results
contrast with earlier work \cite{44} and suggest that more work is
necessary to understand the mechanisms of action and fundamental
limitations of optimal open-loop coherent control in the Markovian
setting.

The situation is very different in the non-Markovian setting.  In this
case, the control, even if its action is limited to the system, can
fully take advantage of memory effects in the environment to restore
coherence to and decouple the system from the environment, given
sufficient time and control strength.  In contrast to the Markovian
case, longer gate operation times are actually advantageous and accurate
knowledge of the system-bath coupling is crucial.  While in the
Markovian case knowledge of the effective strength and uniformity of
environmental effects is sufficient (and there is no need to know the
precise type of coupling), fields optimised without taking the
system-bath coupling into account, or optimised for a different type of
bath, tend to perform poorly in the non-Markovian setting.  This is
consistent with earlier observations in the literature and suggests the
need for accurate characterisation of the system-bath coupling. Further
work is necessary to precisely quantify the degree of knowledge that is
required.  The Bloch trajectory analysis of a subset of the simulations
further suggests that the fields optimised taking the system-bath
coupling into account mainly act by manipulating the system as needed
while at the same time decoupling it from and thus preventing excitation
of the bath.  Specific to the non-Markovian model, field leakage proved
to be a major problem: the final fidelities decreased significantly when
the field also acted on the noise qubits and optimal fields obtained
without considering this effect proved effectively useless.  More
importantly, to restore high fidelities in the presence of field leakage
generally required substantial increases in the gate operation times
unless the frequencies of the noise qubits differed significantly from
those of the system qubits.  The mechanisms of action of the controls
also shifted from suppression of bath excitation --- now no longer
possible --- to exploiting interference effects to effectively
disentangle the system and bath at the target time, while permitting
substantial entanglement between system and bath at intermediate times.
This suggests that field leakage is an important issue in practice that
requires further study.

\ack

SGS acknowledges financial support from EPSRC Advanced Research
Fellowship EP/D07192X/1 and Hitachi.  PdF was supported by a
Hitachi/EPSRC CASE studentship.

\end{document}